%% file: main.tex

\documentclass[11pt]{article}
\usepackage{arxiv}
\usepackage{geometry}                % See geometry.pdf to learn the layout options. There are lots.
\usepackage{graphicx,bm}
\usepackage{epstopdf, epsfig}
\usepackage{xcolor,color}
\usepackage{amsmath}
\usepackage{mathtools}
\usepackage{amssymb}
\usepackage{amsthm}
\usepackage{hyperref}
\usepackage{cleveref}
\usepackage[color=green!20]{todonotes}

\newcommand{\bit}[0]{\textit{b}}
\newcommand{\cit}[0]{\textit{c}}

\newcommand{\bnabla}[0]{\boldsymbol{\nabla}}

\title{Transport Phenomena in Fluid Films with Curvature Elasticity}
\author{Arijit Mahapatra,
David Saintillan,
and Padmini Rangamani \\
Department of Mechanical and Aerospace Engineering, University of California San Diego,\\
9500 Gilman Drive, La Jolla, CA 92093, U.S.A.\\
Email address of corresponding author: \href{prangamani@ucsd.edu}{\color{blue}{prangamani@ucsd.edu}}}

\begin{document}

\maketitle

\begin{abstract}
 Cellular membranes are elastic lipid bilayers that contain a variety of proteins, including ion channels, receptors, and scaffolding proteins. These proteins are known to diffuse in the plane of the membrane and to influence the bending of the membrane. Experiments have shown that lipid flow in the plane of the membrane is closely coupled with the diffusion of proteins. Thus, there is a need for a comprehensive framework that accounts for the interplay between these processes. Here, we present a theory for the coupled in-plane viscous flow of lipids, diffusion of transmembrane proteins, and elastic deformation of lipid bilayers. The proteins in the membrane are modeled such that they influence membrane bending by inducing a spontaneous curvature. We formulate the free energy of the membrane with a Helfrich-like curvature elastic energy density function modified to account for the chemical potential energy of proteins. We derive the conservation laws and equations of motion for this system.  Finally, we present results from dimensional analysis and numerical simulations and demonstrate the effect of coupled transport processes in governing the dynamics of membrane bending and protein diffusion. 
\end{abstract}

\section{Introduction}
%\todods[inline]{David, this is an example of an inline comment from you}
%\todopr[inline]{And my comments will be in pink}
%\todopr[inline]{Outline of introduction: paragraph 1 Cell membranes have proteins and lipids flow in the plane of the membrane. This is known in cells and reconstituted experiments. This is in addition to bending. paragraph 2 Theoretical models have been developed to study these different aspects. summarize key findings.paragraph 3 In this work, we set out to do what?}

Lipid bilayers are present both in the plasma membrane and in intracellular organelles \cite{alberts02} and have an extremely heterogeneous composition \cite{Harayama18}. 
They consist of many different types of lipids, integral, and peripheral membrane proteins \cite{Bassereau18}, all of which are important in cellular function \cite{Singer74}. %\needref.
%The embedded proteins in the lipid bilayer play a pivotal role to govern the morphology of the membrane (\cite{Kim98}). 
One of the classic features of cellular membranes is their ability to bend out of plane and this has been the focus of many studies, both theoretical and experimental, over the past five decades \cite{Sackmann86, Lipowsky91, Julicher93, Zimmerberg06, Hassinger17}. %\needref.
We now also know that these membrane-protein interactions in cells are associated with many curvature sensing \cite{Antonny11} and curvature generating phenomena \cite{McMahon05} including tubulation \cite{Stachowiak10}, vesicle generation \cite{Reynwar07}, and membrane trafficking \cite{Mukherjee00, Gruenberg01}. %\needref. 
Curvature is a shape variable of the membrane that is related to the internal parameters such as protein density, tilt angle, local composition \cite{Leibler87}, and intermonolayer differences of the membrane \cite{seifert1993curvature}.
Proteins embedded in the membrane diffuse in the plane of the membrane and undergo transport by advection processes \cite{Kahya04} associated with viscous flow of the lipids \cite{Tran-son-tay84, Noguchi04}. 
%Indeed, advances in imaging modalities have identified that many transmembrane proteins have a non-negligible diffusion coefficient (\cite{Thomas15}). 
%These dynamics of cellular membranes are critical to cellular function, lipid rafts, and even in governing cell death (\cite{Zhang18}).  
Experimental observations in reconstituted or synthetic lipid vesicles show that the coupling of lipid flow, protein diffusion, and membrane bending can give rise to emergent phenomena \cite{Baumgart03, Horner09, Stachowiak17}. %\todopr{Arijit, the references need to be both complete and current. I'm pretty sure that in addition to this '09 paper, there are more recent papers that talk about interaction between bending and membrane protein interactions.} 
There are theoretical models that have explicitly studied the coupling between concentration of curvature-inducing proteins and the bending of the membrane \cite{Lowengrub2009, Elliot2013, Elliot2016} and coupling between viscous flow and bending \cite{Seifert93b}. The coupling between flow, diffusion and bending has not been commonly considered with the exception of a few phase transition models \cite{sahu2017irreversible}. Thus, there is a need to understand how the interplay of protein diffusion, lipid flow, and membrane bending determines the mechanical response of lipid bilayers.

%Many groups have focused on the development of theoretical models of lipid bilayer mechanics \cite{Seifert97, das11, Alimohamadi18,omar2019non}. %\needref.
The seminal work of \cite{Helfrich73}, \cite{Canham70}, and \cite{Jenkins77} established the framework for using variational principles and thin shell mechanics for modeling membrane bending.
Later, \cite{Steigmann99} established the correspondence between Koiter's shell theory and developed a complete theoretical framework of membrane mechanics. %\todopr{I will tighten this sentence. It's not very clean.}
These early models assumed the membrane to be inviscid and focused primarily on elastic effects.
In the past decade, many groups have proposed the addition of viscous effects in addition to membrane bending \cite{Arroyo09,Arroyo12,Rangamani13,Arroyo19} %\todopr{check references. Also reference the 2019 Arroyo papers. There are two Arroyo09s, which is quite confusing. Which is the Rahimi paper?}
building on the ideas proposed by \cite{Scriven60}.
We also showed recently that including intrasurface viscosity in addition to membrane bending allows for the calculation of local membrane tension in the presence of protein-induced spontaneous curvature \cite{Rangamani14} and for the calculation of flow fields on minimal surfaces \cite{Bahmani15}.
%which derives the governing equation for Stokes flow in a fluid interface. The constitutive relation for stress-strain, used in that model, is taken from \cite{Oldroyd55}.  We also find a well-established methodology for the viscous fluid flow in a Riemannian manifold in the textbook of fluid mechanics (\cite{Aris89}). More recently, we find viscous flow in an elastic lipid bilayer was studied by \cite{Rangamani13}, and \cite{Arroyo09},  and \cite{Arroyo12}. 
Separately, the interaction between in-plane protein diffusion and membrane bending has been modeled \cite{Iglic05,Iglic05a,Gov07,Ramaswamy99,Gozdz11,Reynwar07}. %\needref. \todopr{Arijit, Nir Gov, Iglic and all the relevant references from Haleh's review and Haleh's nanotube paper!}
Specifically, \cite{Agrawal11} proposed a framework that included the chemical potential energy of membrane-protein interactions and membrane bending and demonstrated the interaction between bending and diffusion. 
A series of studies by Arroyo and coworkers also developed a comprehensive framework for incorporating membrane-protein interactions using Onsager's variational principles \cite{Arroyo18,Arroyo19, Arroyo19jfm}. %\needref.

%In addition to these, physicists have studied diffusion in a Riemannian manifold and developed models with the help of constitutive relations. The problem becomes much involved when the manifold (lipid bilayer) bends due to the species (proteins) that diffuse in it. We performed a theoretical study to find the stability of a fluctuating membrane that deforms due to the asymmetry of the floating protein. They have formed a conservation equation to find the protein concentration in a deforming surface by minimizing the total energy.  \cite{Seifert05} studied similar phenomena from the Smolouschovski equation and observed that diffusion of protein is considerably dependent on thermal fluctuation, membrane tension, and bending rigidity. \cite{Gozdz11} studied the shape transformation of lipid vesicles induced by diffusing proteins.  More recently, \cite{Klaus16} observed that curvature in a tubular structure works as a barrier to the diffusion. \cite{Desernoagr} noticed through their molecular dynamics simulation that curvature induces aggregation of protein in the membrane. Furthermore, the presence of flow can influence both the diffusion and bending under convective transport of the protein. This makes necessary to build another comprehensive framework for the diffusion of protein in the viscous membrane coupled with bending. \cite{Arroyo19} studied such a problem in an axisymmetric domain and focus mainly on the entropic aggregation of protein in a vesicle and found the shapes corresponding to that. 

Building on these efforts, we present a coupled theory for membrane mechanics that accounts for in-plane viscous flows and diffusion of curvature-inducing transmembrane proteins in addition to membrane bending. 
We note that a version of this model was presented by \cite{Steigmann2018}.
Using a free energy functional that includes bending energy, chemical potential energy of membrane-protein interactions, and by including the viscous stresses in the force balance, we derive the governing equations of motion in \S\,\ref{sec:Model Development}.
In \S\,\ref{sec:linear_def}, we analyze this system of equations assuming small deformations from the flat plane and identify the role of different dimensionless groups in governing the regimes of operation.
We then perform numerical simulations in a one-dimensional model in \S\,\ref{sec:1ddomain} and in a two-dimensional Monge parametrization in \S\,\ref{sec:2ddomain}. The case of large deformations is addressed in \S\,\ref{sec:axisymmetry} where we investigate the flattening of a membrane bud in axisymmetric coordinates.
These results are cast in perspective of the current knowledge of the field and future directions are presented in \S\,\ref{sec:conclusions}. 

\section{Membranes with intra-surface viscosity and protein diffusion}
\label{sec:Model Development}

We formulate the governing equations for the dynamics of an elastic lipid membrane with surface flow, coupled to the transport of membrane-embedded proteins that induce spontaneous mean curvature. The notations used in the model are summarized in \Cref{table:notation1}. We assume familiarity with
tensor analysis and curvilinear coordinate systems \cite{Sokolnikoff1964,Kreyszig1968,Aris89}.

\subsection{Membrane geometry, kinematics and incompressibility}

The lipid membrane is idealized as a two-dimensional manifold $\Omega$ in three-dimensional space. Material points on $\Omega$  are parametrized by a position field $\boldsymbol{r}(\theta^\alpha,t)$, where $\theta
^{\alpha}$ are surface coordinates and play a role analogous to that of a fixed coordinate system used to parametrize a control volume in the Eulerian description of classical fluid mechanics. Here and henceforth, Greek indices range over $\{1,2\}$
and, if repeated, are summed over that range. %We assume familiarity with
%tensor analysis and curvilinear coordinate systems \cite{Sokolnikoff1964,Kreyszig1968}.
The local tangent basis on the surface is naturally obtained as $\boldsymbol{a}_{\alpha }=\boldsymbol{r}_{,\alpha }$\, where commas identify partial derivatives with respect to $\theta ^{\alpha }.$ The unit normal field is then given by  $\smash{\boldsymbol{n}=\boldsymbol{a}_{1}\times \boldsymbol{a}_{2}/\left\vert \boldsymbol{a}_{1}\times \boldsymbol{a}
_{2}\right\vert}$. The tangent basis also defines the surface metric $a_{\alpha \beta }=\boldsymbol{a}_{\alpha }\boldsymbol{\cdot} 
\boldsymbol{a}_{\beta }$ (or coefficients of the first fundamental form), a positive definite matrix, which is one of the two basic variables in surface theory. The other is the  curvature $b_{\alpha \beta }$ (or
coefficients of the second fundamental form) defined as $b_{\alpha \beta }=%
\boldsymbol{n}\boldsymbol{\cdot} \boldsymbol{r}_{,\alpha \beta }$. Of special interest are the mean and Gaussian curvatures, which will enter the Helfrich energy of the membrane and are defined, respectively, as
\begin{equation}
H=\frac{1}{2}a^{\alpha \beta }b_{\alpha \beta }, \quad
K=\frac{1}{2}\varepsilon ^{\alpha \beta }\varepsilon ^{\mu\eta }b_{\alpha
\mu }b_{\beta \eta }\,.
\label{eqn:curvature}
\end{equation}
Here, $a^{\alpha \beta}=(a_{\alpha \beta })^{-1}$ is the dual metric, and $%
\varepsilon ^{\alpha \beta }$ is the permutation tensor defined as $%
\varepsilon ^{12}=-\varepsilon ^{21}=1/\sqrt{a},$ $\varepsilon
^{11}=\varepsilon ^{22}=0.$ 

We assume that the surface $\Omega$ is moving with time, and the velocity of a material point in the membrane is given by $\boldsymbol{u}(\theta^\alpha,t)=\dot{\boldsymbol{r}}=\partial \boldsymbol{{r}}/\partial t$. It can be expressed in components on the natural basis introduced above:
\begin{equation}
\boldsymbol{u=\,}\,v^{\alpha }\boldsymbol{a}_{\alpha }+w\boldsymbol{n}\,,
\end{equation}%
where the components $v^\alpha$ capture the tangential lipid flow and $w$ is the normal surface velocity. The membrane is  assumed to be incompressible, which prescribes a relationship between the in-plane velocity field and the curvature as \cite{Arroyo09,Rangamani13}
\begin{equation}
v^\alpha_{;\alpha}=2Hw ,
\label{eqn:constraint}
\end{equation}
where the semi-colon refers to covariant differentiation with respect to the metric $a_{\alpha \beta}$.

\subsection{Stress balance and equations of motion}

We model the membrane as a thin elastic shell and, in the absence of inertia, the equations of motion are the equations of mechanical equilibrium. For a membrane subjected to a lateral pressure difference $p$ in the direction of the unit normal $\boldsymbol{n}$, these may be summarized  as \cite{Steigmann99}
\begin{equation}
\boldsymbol{T}_{;\alpha }^{\alpha }+p\boldsymbol{n}=\boldsymbol{0}\,,
\label{eqn:eqm}
\end{equation}%
where $\boldsymbol{T}^{\alpha }$ are the so-called \textit{stress vectors}. The differential operation in equation (\ref{eqn:eqm}) is the surface divergence
\begin{equation}
\boldsymbol{T}_{;\alpha }^{\alpha }=(\sqrt{a})^{-1}(\sqrt{a}\,\boldsymbol{T}^{\alpha
})_{,\alpha }\,,
\label{eqn:stress_balance_1}
\end{equation}%
where $a=\det (a_{\alpha \beta })>0$. 
This framework encompasses all elastic surfaces for which the energy density
responds to metric and curvature. For example, if the energy density per
unit mass of the surface is $F(a_{\alpha \beta },b_{\alpha \beta }),$ then \cite{Steigmann99}
\begin{equation}
\boldsymbol{T}^{\alpha }=N^{\beta \alpha }\boldsymbol{a}_{\beta }+S^{\alpha }\boldsymbol{n},
\label{eqn:stress_balance_2}
\end{equation}%
where
\begin{equation}
 N^{\beta \alpha }=\zeta^{\beta \alpha }+\pi^{\beta\alpha}+b_{\mu }^{\beta }M^{\mu
\alpha} \qquad \mathrm{and} \qquad S^{\alpha }=-M_{;\beta }^{\alpha \beta }.
\label{eqn:stresses}
\end{equation}
We have introduced the notation $\smash{b_{\beta }^{\alpha }=a^{\alpha \lambda }b_{\lambda \beta}}$. In equation (\ref{eqn:stresses}), $\smash{\zeta^{\beta\alpha}}$ is the in-plane elastic stress tensor, $\smash{\pi^{\beta\alpha}}$ is the intra-membrane viscous stress tensor due to surface flow, and $M^{\alpha \beta}$ is the moment tensor due to curvature-induced elastic bending. We discuss constitutive equations for these various contributions in the next sections.  

Substituting equations (\ref{eqn:stress_balance_2}) and (\ref{eqn:stresses}) into (\ref{eqn:eqm}), invoking the Gauss and Weingarten
equations $\boldsymbol{a}_{\beta ;\alpha }=b_{\beta \alpha }%
\boldsymbol{n}$ and $\boldsymbol{n}_{,\alpha }=-b_{\alpha }^{\beta }\boldsymbol{a}%
_{\beta }$ \cite{Sokolnikoff1964}, and projecting the result onto the tangent and normal spaces of $\Omega$ provides the three governing equations
\refstepcounter{equation}
$$
N_{;\alpha }^{\beta \alpha }-S^{\alpha }b_{\alpha }^{\beta }=0, \qquad S_{;\alpha }^{\alpha }+N^{\beta \alpha }b_{\beta \alpha }+p=0\,, \eqno{(\theequation{\mathit{a}, \mathit{b}})}. \label{eqn:eqns_of_motion}
$$
which express stress balances in the tangential and normal directions.

\subsection{Free energy of an elastic membrane with curvature-inducing proteins}

The elastic contribution of the surface stress and the moment tensor are derived from a free energy and are expressed as \cite{Steigmann99} 
\begin{equation}
\zeta ^{\beta \alpha}=\rho \left(\frac{\partial F}{\partial a_{\alpha \beta }}+ \frac{\partial F}{\partial a_{\beta \alpha }}\right), \qquad  M^{\alpha \beta}=\frac{\rho}{2} \left(\frac{\partial F}{\partial b_{\alpha \beta}}+\frac{\partial F}{\partial b_{\beta \alpha}} \right).
\label{eqn:elastic_sigma}
\end{equation}
Here, $F$ is the energy Lagrangian per unit mass defined as \cite{Steigmann99}
\begin{equation}
    F(H,K,\rho)=\bar{F}(H,K)-\gamma/\rho,
\end{equation}
where $\gamma$ is a Lagrange multiplier imposing the constraint of incompressibility, and $\rho$ is the membrane density which is assumed to be constant. It is customary to formulate the mechanics in terms of the free energy per unit area as \cite{Steigmann03}
\begin{equation}
W=\rho \bar{F}.
\end{equation}
For an elastic membrane with a density $\sigma$ of curvature-inducing proteins, we model this free energy as the sum of elastic and chemical energies \cite{Gov07, Agrawal11,Alimohamadi18} %needref
\begin{equation}
%W(H,K,\sigma )=k[H-C(\sigma )]^{2}+\bar{k}K+\mathcal{A} (\sigma).  \label{eq:freeenergy}
W(H,K,\sigma )=k[H-C(\sigma )]^{2}+\bar{k}K+k_BT\,\sigma\Big[\log\Big(\frac{\sigma}{\sigma_s}\Big)-1\Big].  \label{eq:freeenergy}
\end{equation}%
The first two terms correspond to the classical Helfrich free energy and involve the two bending moduli $k$ and $\bar{k}$. While these could in general depend on $\sigma$, we take them to be constant as is appropriate in the dilute limit. $C(\sigma)$ is the protein-induced spontaneous curvature and is assumed to depend linearly on protein density  \cite{Gov07, Agrawal11,Gov2018}:
\begin{equation}
C=\ell\sigma\,,
\end{equation}
where the constant $\ell$ is a characteristic length scale associated with the embedded protein.
The last term in  equation (\ref{eq:freeenergy}) is the entropic contribution due to thermal diffusion of proteins \cite{Gov2018}, where $k_BT$ is the thermal energy and $\sigma_s$ denotes the saturation density of proteins on the membrane. 

%The last term in  (\Cref{eq:freeenergy}) is the chemical energy of membrane-protein interaction. In dilute solutions, the chemical potential $\chi$ varies linearly with the protein density $\sigma$ \cite{Landau59} and as a result $\mathcal{A}$ is at least a quadratic function of $\sigma$. Here, we use the form $\mathcal{A} (\sigma)=(A_1\sigma+A_0)^2$ as proposed by \cite{Agrawal11}, where $A_0$ and $A_1$ are positive constants. 

Inserting equation~(\ref{eq:freeenergy}) for the free energy into equation~(\ref{eqn:elastic_sigma}) provides expressions for the elastic stress and moment tensors as
\begin{align}
\zeta^{\alpha\beta}&=-2k(H-\ell \sigma) b^{\alpha \beta}-2\bar{k}Ka^{\alpha \beta}-\gamma a^{\alpha \beta},  \label{eq:zeta} \\
M^{\alpha\beta}&=k(H-\ell \sigma) a^{\alpha\beta}+\bar{k}(2Ha^{\alpha\beta}-b^{\alpha\beta}).\label{eq:Mab}
\end{align}

\subsection{Viscous stress}

 In the presence of surface flow, a viscous stress also develops in the membrane. The deviatoric part of the viscous stress tensor $\pi ^{\alpha \beta }$ is assumed to depend linearly on strain rate as \cite{Scriven60}
\begin{equation}
\pi ^{\alpha \beta }=\nu a^{\alpha \mu }a^{\beta \eta }\dot{a}_{\mu \eta }, 
\end{equation}%
where $\nu$, a positive constant, is the intra-membrane surface viscosity.
Further expanding this expression, we obtain
\begin{equation}
\pi ^{\alpha \beta }=2\nu \lbrack a^{\alpha \mu }a^{\beta \eta }d_{\mu\eta}-wb^{\alpha \beta }],
\label{eqn:viscous_stress}
\end{equation}
where $d_{\mu\eta }=(v_{\mu ;\eta }+v_{\eta ;\mu })/2$ is the surface rate-of-strain tensor.

\subsection{Conservation equation for protein transport}

To complete the model, we specify a transport equation for the protein density $\sigma$. Mass conservation can be expressed as
\begin{equation}
\label{eqn:diff}
\frac{\partial \sigma}{\partial t} +m^{\alpha}_{;\alpha}=0\,,  
\end{equation}
where $m^\alpha$ denotes the protein flux. This flux has contributions from advection by the surface flow as well as from gradients in chemical potential. Following \cite{Agrawal11}, it can be derived from first principles as
\begin{equation}
    m^\alpha=\Big[v^\alpha -\frac{1}{f}a^{\alpha\beta}W_{\sigma,\beta}\Big]\sigma\,,  
    %m^\alpha=-c\,a^{\alpha \beta}W_{\sigma,\beta}+{v}^\alpha\sigma\,,
\end{equation}
where $f$ is the hydrodynamic drag coefficient of a protein and $W_\sigma=\partial W/\partial\sigma$. 

\subsection{Summary of the governing equations\label{sec:gov_eqn}}

We summarize the governing equations for the membrane and protein dynamics. The tangential momentum balance, obtained by inserting equations (\ref{eq:zeta}), (\ref{eq:Mab}) and (\ref{eqn:viscous_stress}) for the stresses into the first equation in (\ref{eqn:eqns_of_motion}), is expressed as 
\begin{equation}
    \lambda _{,\alpha }-4\nu wH_{,\alpha }+2\nu (a^{\beta\mu }d_{\alpha \mu
;\beta }-w_{,\beta }b_{\alpha }^{\beta })=-\sigma_{,\alpha} \Big[ k_BT\log\Big(\frac{\sigma}{\sigma_s}\Big) -2k \ell (H-\ell \sigma) \Big],
%  \lambda _{,\alpha }-4\nu wH_{,\alpha }+2\nu (a^{\beta\mu }d_{\alpha \mu
%;\beta }-w_{,\beta }b_{\alpha }^{\beta })=-\sigma_{,\alpha} \left[ 2A_1 (A_1 \sigma +A_0)^2 -2k \ell (H-\ell \sigma) \right], 
\label{eq:stokes}
\end{equation}
where we have introduced the membrane tension $\lambda=-(W+\gamma)$. Along with the surface incompressibility condition  
\begin{equation}
    v_{;\alpha }^{\alpha}-2wH=0,
\end{equation}
equation~(\ref{eq:stokes}) constitutes the governing equation for the intra-membrane flow. We note the similarity with the Stokes equations, where the tension $\lambda$ plays a role analogous to the pressure in classical incompressible flow \cite{Bahmani15}. The right-hand side captures the forcing by the protein distribution on the flow. Similarly, the normal force balance in (\ref{eqn:eqns_of_motion}) provides the so-called \textit{shape equation}, written after simplifications as 
\begin{align}
%\begin{split}
k\Delta (H-\ell\sigma)+2k(H-\ell\sigma)(2H^{2}-&K)  -2H\Big[k_BT\sigma\Big(\log\Big(\frac{\sigma}{\sigma_s}\Big)-1\Big)+k(H-\ell\sigma)^{2}\Big]\nonumber \\
&- 2\nu \big[ b^{\alpha \beta }d_{\alpha
\beta } -w(4H^{2}-2K)\big] =p+2\lambda H,
%k\Delta (H-\ell\sigma)+2k(H-\ell\sigma)(2H^{2}-K)-2H[(A_1 \sigma + A_0
%)^{2}+k(H-\ell\sigma)^{2}] + \\
%2\nu \lbrack b^{\alpha \beta }d_{\alpha
%\beta } -w(4H^{2}-2K)] =p+2\lambda H,
%\end{split}
\label{eq:shapeeq}
\end{align}
where $\Delta(\cdot)=(\cdot)_{;\alpha\beta}a^{\alpha\beta}$ is the surface Laplacian. Equation (\ref{eq:shapeeq}) can be interpreted as the governing equation for the position field $\boldsymbol{r}$. Finally, the model is completed by the advection-diffusion equation for the protein density, which is written
%\begin{equation}
%    \frac{\partial \sigma}{\partial t}+ (v^{\alpha} \sigma)_{;\alpha} = (D +2ck\ell^2) a^{\alpha \beta} \sigma_{,\alpha \beta} -2c k \ell a^{\alpha \beta} H_{,\alpha \beta}.
%\end{equation}
\begin{equation}
\label{eqn:adv_diff}
    \frac{\partial \sigma}{\partial t}+ (v^{\alpha} \sigma)_{;\alpha} = Da^{\alpha \beta}\sigma_{,\alpha \beta}-\frac{2k \ell }{f}\big[a^{\alpha \beta}(H-\ell \sigma)_{,\alpha} \sigma_{,\beta}+\sigma a^{\alpha \beta}(H-\ell \sigma)_{,\alpha \beta}\big].
   % \,a^{\alpha \beta} \sigma_{,\alpha \beta}  -2ck\ell a^{\alpha \beta} ( H_{,\alpha \beta} - \ell \sigma_{,\alpha \beta}) .
    %\frac{\partial \sigma}{\partial t}+ (v^{\alpha} \sigma)_{;\alpha} = D\,a^{\alpha \beta} \sigma_{,\alpha \beta}  -2ck\ell a^{\alpha \beta} ( H_{,\alpha \beta} - \ell \sigma_{,\alpha \beta}) .
\end{equation}
%\todods{To be consistent, write the right-hand side of (2.23) in index notation}
%\todopr{The equation is correct as written. However, I agree with David that it will lead to the confusion that curvature can create protein. I think we should discuss this equation in the context of the boundary conditions and how it differs from Fickian diffusion. I'll do that later today after I read couple of papers.}
The first term on the right-hand side captures Fickian diffusion of proteins,  with diffusivity $D$ given by the Stokes-Einstein relation: $D=k_B T/f$. The second term captures the interaction of the curvature and protein gradients. The third term captures the effect of membrane shape on protein transport: mismatch between the mean curvature and the protein-induced  spontaneous curvature serves as a source term for the transport of protein on the membrane surface.
%with departures from the protein-induced spontaneous curvature driving a flow of protein.  
%departures in  sideWe have introduced the diffusion coefficient  $D=2cA_1^2$. The first term on the right-hand side is similar to Fickian diffusion. The second term  except that the net diffusion coefficient has contributions from the protein diffusion coefficient $D$ and contributions from the bending modulus and curvature induced by the proteins. \todods{Is there a better way of saying this?} \todopr{I expanded it, let me know what you think.} The coupling with membrane deformations also results in the second term, capturing the direct effect of mean curvature on protein transport.  

\begin{table} 
%\begin{center}
\centering
\caption{Summary of the notations used in the model.}
\begin{tabular} {l l l}
\hline\hline
Notation &  Description & Units\\ [0.5ex]
\hline
$\gamma$ & Lagrange multiplier for incompressibility   & pN$\,\cdot\,$nm$^{-1}$ \\
$p$ & Pressure difference across the membrane  & pN$\,\cdot\,$nm$^{-2}$ \\
$C$ & Protein-induced spontaneous curvature & nm$^{-1}$ \\
$\theta^{\alpha}$ & Surface coordinates \\ 
$W$ & Local free energy per unit area &  pN$\,\cdot\,$nm$^{-1}$\\
$\lambda$ & Membrane tension, $\lambda=-(W+\gamma)$  & pN$\,\cdot\,$nm$^{-1}$\\
$\lambda_0$ & Membrane tension at infinity  & pN$\,\cdot\,$nm$^{-1}$\\
${\zeta}^{\alpha \beta}$ & Elastic stress tensor   & pN$\,\cdot\,$nm$^{-1}$\\
${\pi}^{\alpha \beta}$ & Viscous deviatoric stress   & pN$\,\cdot\,$nm$^{-1}$\\
$H$ & Mean curvature of the membrane & nm$^{-1}$\\
$K$ & Gaussian curvature of the membrane & nm$^{-2}$\\
$k$ & Bending modulus (rigidity)  & pN$\,\cdot\,$nm\\
$\bar{k}$ & Gaussian modulus & pN$\,\cdot\,$nm\\
$\sigma$ & Protein density per unit area & nm$^{-2}$\\
$\sigma_s$ & Saturation protein density per unit area & nm$^{-2}$\\
%$\sigma_0$ & Initial protein density at the patch & nm$^{-2}$\\
$\ell$ & Proportionality constant of $C$ -- $\sigma$ relation & nm\\
%$\phi$ & Angle made by the protein &  $-$ \\
%$c$ & Diffusion coefficient of protein & nm$^2/$s\\ 
%$\chi$ & Chemical potential & pN$\,\cdot\,$nm \\
%$A_1$ & Slope of linear relation between $\chi$ and  $\sigma$  &  pN$^{1/2} \cdot$ nm$ ^{3/2}$ \\
%$A_0$ & Intercept of linear relation between $\chi$ and  $\sigma$  &  pN$^{1/2}\cdot$ nm $^{-1/2}$ \\
$D$ & Protein diffusion coefficient, $D=k_BT/f$ & nm$^2\cdot\,$s$^{-1}$\\ 
$f$ & Hydrodynamic drag coefficient of a protein &  pN$\cdot$s$\cdot$nm $^{-1}$ \\
$L$ & Size of the domain & nm \\
$k_B$ & Boltzmann constant & pN$\cdot$nm$\cdot$K$^{-1}$ \\
$T$ & Temperature & K \\ 
%$\boldsymbol{c}$ & Velocity vector & nm$/$s \\
%$v_{ref}\left(=\frac{\lambda_0L}{\nu}\right)$ & Reference velocity scale & nm$/$s \\
%$t_{ref}\left(=\frac{L^2}{D}\right)$ & Reference time scale & s \\
\hline
\end{tabular}
%\end{center}
\label{table:notation1}
\end{table}

\subsection{Comparison with other existing models}

The equations of motion (\ref{eqn:eqns_of_motion}) have appeared in the literature in many different forms. For example, the tangential stress balance (\ref{eqn:eqns_of_motion}$a$) is similar to equation (9) of \cite{Julicher17} and equation (1) of \cite{Mietke19}. We eventually obtain (\ref{eq:stokes}) from this equation by using the constitutive relation (\ref{eqn:viscous_stress}) for the viscous stress, an equivalent expression of which can be found in equation (D11) of \cite{Arroyo19jfm}. The final form of the tangential balance (\ref{eq:stokes}) also correlates to their equation (D9). Similarly, the normal force balance relation (\ref{eqn:eqns_of_motion}$b$) compares with equation (10) of \cite{Julicher17} and equation (2) of \cite{Mietke19}. The final form of the force balance relations (\ref{eq:stokes})--(\ref{eq:shapeeq}) captures the effect of curvature-inducing proteins that diffuse in the plane of the membrane. This differs from the model of \cite{Mietke19} where the effect of protein-induced active tension was considered, or from the model of \cite{Julicher17}, which includes a contribution from the external stress and torque applied by the extracellular matrix. %\cite{Arroyo19jfm}, on the other hand, derived all the conservation relation and corresponding ALE formulation followed by time integration of different field distribution on that surface. 
The advection-diffusion equation (\ref{eqn:adv_diff}) in our model is also similar to equation (13) of \cite{Mietke19}, equation (3.27) of \cite{mikucki2017curvature} and equation (4) of \cite{gera2017cahn}. These three studies, however, did not include the strong coupling between bending and diffusion in equation (\ref{eqn:adv_diff}), which results from the curvature inducing property of the membrane proteins. In the model presented by \cite{Mietke19}, the coupling between bending and diffusion occurs through an active tension induced by the proteins. \cite{mikucki2017curvature} calculated the phase-field of protein density  
in an inviscid framework, while \cite{gera2017cahn} solved a Cahn-Hilliard equation on a preexisting curved surface.

\section{Small deformations from the flat plane\label{sec:smalldef}}
\subsection{Linearization and dimensional analysis\label{sec:linear_def}}

In this section, we specialize the governing equations presented in \S\,\ref{sec:gov_eqn} in a Monge parametrization assuming small deflections from the flat plane.

\subsubsection{Governing equations in the linear deformation regime}

The surface parametrization for a Monge patch is given by
\begin{equation}
    \boldsymbol{r}(x,y,t)=x\boldsymbol{i}+y\boldsymbol{j}+z(x,y,t)\boldsymbol{k},
\end{equation}
where unit vectors $(\boldsymbol{i},\boldsymbol{j},\boldsymbol{k})$ form a fixed Cartesian orthonormal basis, and $z(x,y,t)$ is the deflection from the $(x,y)$ plane.
The tangent and normal vectors are given by
\begin{align}
\boldsymbol{a}_1=\boldsymbol{i}+z_{,x}\boldsymbol{k}, \quad \boldsymbol{a}_2=\boldsymbol{j}+z_{,y}\boldsymbol{k}, \quad \boldsymbol{n}=\frac{1}{(1+z_{,x}^2+z_{,y}^2)^{1/2}}(-z_{,x} \boldsymbol{i}-z_{,y} \boldsymbol{j}+\boldsymbol{k}).
\end{align}
The surface metric ($a_{\alpha \beta}$) and curvature metric ($b_{\alpha \beta}$) take the following forms
\begin{equation}
   a_{\alpha \beta}=
  \left[ {\begin{array}{cc}
   1+z_{,x}^2 & z_{,x} z_{,y} \\
   z_{,y} z_{,x} & 1+z_{,y}^2 \\
  \end{array} } \right], \quad
  b_{\alpha \beta}=\frac{1}{(1+z_{,x}^2+z_{,y}^2)^{1/2}}
  \left[ {\begin{array}{cc}
   z_{,xx} & z_{,xy} \\
   z_{,yx} & z_{,yy} \\
  \end{array} } \right].
\end{equation}

We further assume that deflections of the membrane from the flat configuration are small and simplify the governing equations in the limit of weak surface gradients $|\boldsymbol{\nabla} z|\ll 1$ by neglecting quadratic terms in $|\boldsymbol{\nabla} z|$ \cite{Do76}. In this limit, differential operators in the space of the membrane reduce to the Cartesian gradient, divergence and Laplacian in the $(x,y)$ plane. The linearized governing equations for the intra-membrane flow become:
\begin{equation}
\boldsymbol{\nabla}\boldsymbol{\cdot}\boldsymbol{v}=2wH,
\end{equation}
\begin{equation}
\begin{split}
\boldsymbol{\nabla} \lambda +\nu {\nabla}^2 \boldsymbol{v}+\nu \boldsymbol{\nabla}(\boldsymbol{\nabla}\boldsymbol{\cdot}\boldsymbol{v})-4\nu w \boldsymbol{\nabla} H-2\nu \boldsymbol{\nabla} w \boldsymbol{:} \boldsymbol{\nabla}\boldsymbol{\nabla}z & \\
=k \left({\nabla}^2 z-2\ell \sigma \right) \ell\,  \boldsymbol{\nabla} \sigma -   k_BT\log\Big(\frac{\sigma}{\sigma_s}\Big)&\boldsymbol{\nabla} \sigma, 
%2A_1 (A_1 \sigma +A_0)&\boldsymbol{\nabla} \sigma, 
\end{split}
\end{equation}
%\todods{Provide expression of $\boldsymbol{b}$ for Monge}
%\todoam{I have used the expression of $\boldsymbol{b}$ here directly, will this work?}
%\todopr{Arijit, give the expression of both $\boldsymbol{a}$ and $\boldsymbol{b}$ earlier after you define the tangents and normals. This should be after equation 3.2}
whereas the shape equation expressing the normal momentum balance is
\begin{align}
\begin{split}
 k\left(\tfrac{1}{2} {\nabla}^4 z -\ell {\nabla}^2 \sigma \right)-{\nabla}^2 z \Big[k_BT\sigma\Big(\log\Big(\frac{\sigma}{\sigma_s}\Big)-1\Big)+k \ell^2 \sigma^2\Big]&  \\
- \nu \big(\boldsymbol{\nabla} {\boldsymbol{v}}+ \boldsymbol{\nabla}\boldsymbol{v}^T\big)\boldsymbol{:}\boldsymbol{\nabla} \boldsymbol{\nabla}z
=&\,p+\lambda{\nabla}^2 z.  
\end{split}\label{eqn:linearshapend}
\end{align}
The transport equation for the protein density simplifies to
\begin{equation}
\frac{\partial {\sigma}}{\partial t}+ \bnabla\boldsymbol{\cdot}(\sigma \boldsymbol{v})=D{\nabla}^2 \sigma-\frac{k\ell }{f} \bnabla  \left({\nabla}^2 z -2\ell \sigma \right)\boldsymbol{\cdot} \bnabla \sigma-\frac{k\ell \sigma}{f} \left( {\nabla}^4 z -2\ell {\nabla}^2 \sigma \right).
\end{equation}
We also note the linearized kinematic relation for the normal velocity:
\begin{equation}
w=\frac{\partial z}{\partial t}.
\end{equation}

\subsubsection{Non-dimensionalization}
\label{sec:nond_Monge}
We scale this system of equations using the following reference values. Length is non-dimensionalized by the size $L$ of the domain, protein density by its reference value $\sigma_0$, and membrane tension by its far-field value $\lambda_0$. We also use the characteristic velocity scale $v_c=\lambda_0 L/\nu$ and time scale $t_c= L^2/D$. Denoting dimensionless variables with a tilde, the scaled governing equations are:
\begin{align}
\label{eqn:nondgov_cont}
\tilde{\boldsymbol{\nabla}}\boldsymbol{\cdot}\tilde{\boldsymbol{v}}=2\tilde{w} \tilde{H},\qquad\qquad\quad \\[4pt]
\begin{split}
\label{eqn:nondgov_tang}
\tilde{\boldsymbol{\nabla}} \tilde{\lambda} +\tilde{\nabla}^2 \tilde{\boldsymbol{v}}+\tilde{\boldsymbol{\nabla}} (\tilde{\boldsymbol{\nabla}}\boldsymbol{\cdot} \tilde{\boldsymbol{v}})-4\tilde{w}\tilde{\boldsymbol{\nabla}}\tilde{H}-2\tilde{\boldsymbol{\nabla}}\tilde{w}\boldsymbol{:}\tilde{\boldsymbol{\nabla}}\tilde{\boldsymbol{\nabla}}\tilde{z}&=\\
\tilde{\boldsymbol{\nabla}} \tilde{\sigma} \bigg[\frac{{\vphantom{A}\smash{2 \hat{C} \hat{B}}}}{\hat{T}} \tilde{\nabla}^2 \tilde{z} -\frac{4 \hat{C}^2 \hat{B}^2}{\hat{T}}\, \tilde{\sigma} \,-&\,\frac{2\hat{C}}{\hat{T}} \log\Big(\frac{\tilde{\sigma}}{\tilde{\sigma}_s} \Big) \bigg],
\end{split} \\[4pt]
\begin{split}
\label{eqn:nondgov_norn}
\tilde{\nabla}^4\tilde{z}-{\vphantom{A}\smash{2 \hat{C} \hat{B}}}\,\tilde{\nabla}^2 \tilde{\sigma}  - \hat{C} \tilde{\nabla}^2 \tilde{z}\,\bigg[ 2\tilde{\sigma}\bigg(\log \Big(\frac{\tilde{\sigma}}{\tilde{\sigma}_s}\Big)-1 \bigg)+2 \hat{C} \hat{B}^2 \tilde{\sigma}^2\bigg]&\\
 -\, \hat{T} (\tilde{\boldsymbol{\nabla}} \tilde{\boldsymbol{v}}+\tilde{\boldsymbol{\nabla}} \tilde{\boldsymbol{v}}^T)\boldsymbol{:}\tilde{\boldsymbol{\nabla}}  \tilde{\boldsymbol{\nabla}}\tilde{z} =&\,\hat{P}+\hat{T} \,\tilde{\lambda}\, \tilde{\nabla}^2 \tilde{z},
 \end{split} \\[7pt]
 \label{eqn:nondgov_diff}
 \begin{split}
 \smash{\frac{\partial \tilde{\sigma}}{\partial \tilde{t}}}+Pe \big(\tilde{\boldsymbol{v}}\boldsymbol{\cdot} \tilde{\boldsymbol{\nabla}} \tilde{\sigma}+\tilde{\sigma}\,\tilde{\boldsymbol{\nabla}}\boldsymbol{\cdot}\tilde{\boldsymbol{v}}\big)=\big(1+2 \hat{C} \hat{B}^2\tilde{\sigma}\big)\tilde{\nabla}^2 \tilde{\sigma}\\[2pt]
 + 2 \hat{C} \hat{B}^2 |\tilde{\boldsymbol{\nabla}}\tilde{\sigma}|^2-\,\hat{B}\big(\tilde{\sigma}\tilde{\nabla}^4\tilde{z}&+\tilde{\boldsymbol{\nabla}}\tilde{\nabla}^2\tilde{z}\boldsymbol{\cdot} \tilde{\boldsymbol{\nabla}} \tilde{\sigma}\big).
 \end{split}
\end{align}
The expression for the normal velocity also becomes:
\begin{equation}
\label{eqn:kin_nondw}
    \tilde{w}=\frac{1}{Pe}\frac{\partial \tilde{z}}{\partial \tilde{t}}.
\end{equation}
The dynamics are governed by five dimensionless parameters defined as follows. 
The ratio of the chemical potential to the bending rigidity of the membrane is denoted by $\hat{C}={L^2 k_B T\sigma_0}/{k}$. The ratio of the length scale induced by the proteins and the membrane domain is given by $\hat{L}=\ell L \sigma_0$.
The ratio of the intrinsic length scale of the membrane to the domain size is given by $\hat{T}={2L^2 \lambda_0}/{k}$.
The ratio between the bulk pressure and  bending rigidity is denoted by $\hat{P}={2L^3p}/{k}$. Finally, the P\'eclet number $Pe={\lambda_0 L^2}/{ \nu D}$ compares the advective transport rate to the diffusive transport rate.
We define $\hat{B}=\hat{L}/\hat{C}$ for convenience of simulations and cast the equations in terms of $\hat{L}$, $\hat{B}$, $\hat{T}$, and $Pe$. Further, we assume that there is no pressure difference across the membrane ($\hat{P}=0$).
%The dimensionless chemical potential $C_h={A_0}/{A_1 \sigma_0}$ represents the ratio between the reference chemical potential of the membrane and the protein-induced chemical potential. 
  %leaving four dimensionless groups in the problem. 

\subsection{One-dimensional simulations}
\label{sec:1ddomain}

We first explore the interplay between membrane bending and protein diffusion in the special case of a membrane that deforms as a string in one dimension, with a shape parameterized as $\tilde{z}(\tilde{x},\tilde{t})$.
The flow of lipids does not play a role in this scenario, and as a result in-plane velocity-dependent terms vanish in equations (\ref{eqn:nondgov_cont})--(\ref{eqn:nondgov_diff}). 
The system of governing equations reduces to 
\begin{align}
 \frac{\partial \tilde{\lambda}}{\partial \tilde{x} }= \frac{\partial \tilde{\sigma} }{\partial \tilde{x} }\bigg[\frac{{\vphantom{A}\smash{2 \hat{C} \hat{B}}}}{\hat{T}} \frac{\partial^2 \tilde{z}}{\partial \tilde{x}^2} -\frac{4 \hat{C}^2 \hat{B}^2} {\hat{T}} \tilde{\sigma} -\frac{2 \hat{C}}{\hat{T}}\log\Big(\frac{\tilde{\sigma}}{\tilde{\sigma}_s}\Big)\bigg],     \label{eqn:1dstressbalance}\\
\frac{\partial \tilde{\sigma}}{\partial \tilde{t}}=(1+2 \hat{C} \hat{B}^2 \tilde{\sigma})\frac{\partial^2\tilde{\sigma}}{{\partial {\tilde{x} }}^2 } + 2 \hat{C} \hat{B}^2 \Big( \frac{\partial \tilde{\sigma}}{\partial \tilde{x}}\Big)^2-\hat{B}\left[\tilde{\sigma}\frac{\partial^4 \tilde{z}}{{\partial {\tilde{x} }}^4 }+\frac{\partial \tilde{\sigma}}{\partial \tilde{x}}\frac{\partial }{\partial \tilde{x}}\Big( \frac{\partial^2 \tilde{z}}{\partial \tilde{x}^2} \Big)\right],\\
\frac{\partial^4 \tilde{z}}{{\partial {\tilde{x}  }} ^4}-{\vphantom{A}\smash{2 \hat{C} \hat{B}}} \frac{\partial^2 \tilde{\sigma}}{{\partial {\tilde{x}  }} ^2}  - \hat{C} \frac{\partial^2\tilde{z}}{{\partial {\tilde{x}  }} ^2}\bigg[ 2\tilde{\sigma}\bigg(\log \Big(\frac{\tilde{\sigma}}{\tilde{\sigma}_s}\Big)-1 \bigg)+2 \hat{C} \hat{B}^2 \tilde{\sigma}^2\bigg]
    =\hat{P}+\hat{T} \tilde{\lambda} \frac{\partial^2 \tilde{z}}{{\partial {\tilde{x}  }} ^2}.
    \label{eqn:1dlinearshape}
\end{align}
Equations (\ref{eqn:1dstressbalance})--(\ref{eqn:1dlinearshape}) 
are solved numerically using a finite-difference scheme coded in Fortran 90.\footnote{Numerical codes are available at \\ \hspace*{1.8cm}\url{https://github.com/armahapa/transport_phenomena_in_membranes}} The tangential momentum balance (\ref{eqn:1dstressbalance}), which can be viewed as an equation for the tension $\tilde{\lambda}$, is solved subject to the condition $\tilde{\lambda}(\tilde{x}=1)=1$, whereas the shape equation (\ref{eqn:1dlinearshape}) is solved subject to clamped boundary conditions $\tilde{z}=0$ and $\partial\tilde{z}/\partial \tilde{x}=0$ at both ends of the domain $\tilde{x}=-0.5,0.5$. 

\begin{figure}
    \centering
    \includegraphics[width=\textwidth]{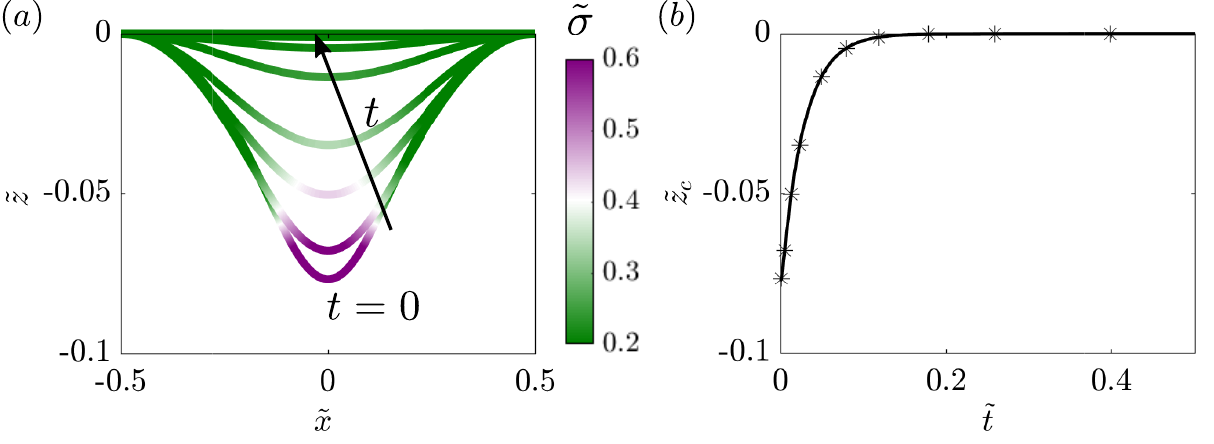}\vspace{-0.2cm}
    \caption{Protein density and membrane deformation in one dimension as functions of time, when an initial protein distribution and no-flux boundary conditions are prescribed.  (\textit{a}) Distribution of protein density plotted on the deformed one-dimensional membrane. (\textit{b}) Time evolution of the maximum membrane deflection $\tilde{z}_c(\tilde{t})=\tilde{z}(1/2,\tilde{t})$.} 
    \label{fig:1dnf}
\end{figure}

We first analyzed the evolution of a symmetric patch of protein defined as $\tilde{\sigma}(\tilde{x},\tilde{t}=0)=1/2[\tanh(20(\tilde{x}+0.1))-\tanh(20 (\tilde{x}-0.1))]$, subject to no-flux boundary conditions on $\tilde{\sigma}$ at the ends of the domain. Results from these simulations are shown in figure~\ref{fig:1dnf}. In response to this protein distribution, the initial configuration of the membrane is bent (see figure~\ref{fig:1dnf}($a$) at $t=0$).
Over time, $\tilde{\sigma}$ homogenizes as a result of diffusion, and therefore the deflection $\tilde{z}$ decreases. 
At steady state, the distribution of protein is uniform on the membrane and $\tilde{z}$ is everywhere zero.
The time evolution of $\tilde{z}$ at the center of the string, corresponding to the maximum deflection, is shown in figure~\ref{fig:1dnf}$(b)$.

%The homogeneous protein distribution as a steady state is easily understood as an effect of diffusion but whether the resulting membrane configuration should be a flat one is unclear when these proteins induce a curvature on the membrane. 

\begin{figure}
    \centering
    \includegraphics[width=1.0\textwidth]{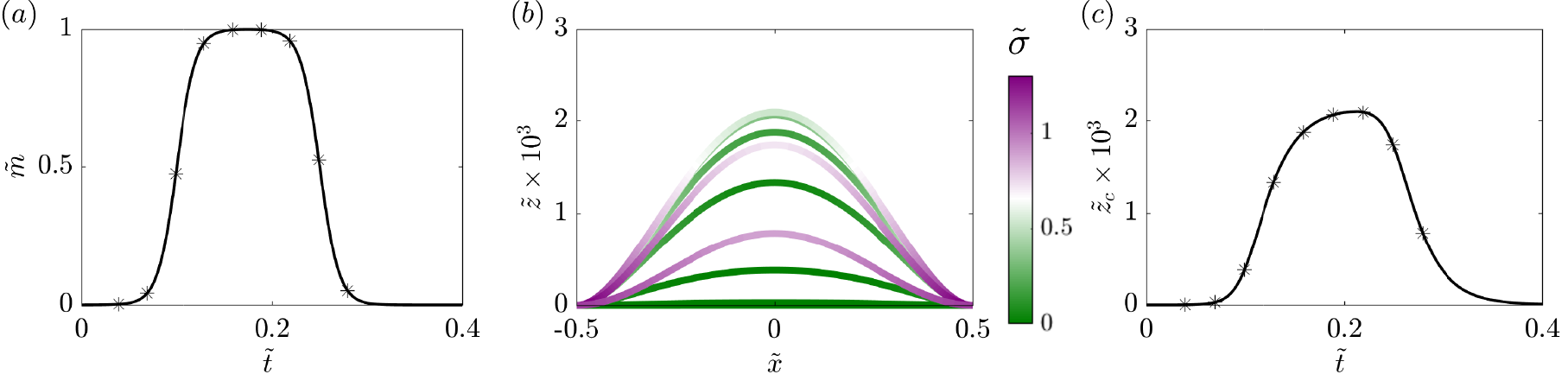}\vspace{-0.2cm}
    \caption{Evolution of membrane deformation and protein distribution when an influx of protein is prescribed at both boundaries.  (\textit{a}) Dimensionless boundary protein flux as a function of time. (\textit{b}) Distribution of protein density plotted on the deformed membrane for $\smash{\hat{C}}=2.48\times10^{-1}$.  (\textit{c}) Time evolution of the maximum membrane deflection $\tilde{z}_c$  for $\smash{\hat{C}}=2.48\times10^{-1}$. Symbols in panels ($a$) and ($c$) correspond to the times shown in (\textit{b}).  }  
    \label{fig:1dflux}
\end{figure}

As a second example, we discuss the case where the protein density is initially zero and a time-dependent protein flux is prescribed at both boundaries as shown in figure~\ref{fig:1dflux}(\textit{a}). In response to the influx at the boundaries, the membrane deforms out of plane as the protein density increases; see figure~\ref{fig:1dflux}($b$). Once the flux returns to zero, diffusion homogenizes the protein, and the membrane height begins to decrease again. This effect is observed clearly by looking at the deformation at the center of the string as a function of time in figure~\ref{fig:1dflux}($c$), which closely follows the dynamics of the boundary flux show in figure~\ref{fig:1dflux}$(a)$.

In both examples of figures~\ref{fig:1dnf} and \ref{fig:1dflux}, we note that the protein distribution becomes uniform at long times (in the absence of any boundary flux), and as a result the membrane returns to its flat reference shape. 
At first glance, this result seems counter-intuitive since there is a non-zero density of curvature-inducing proteins on the membrane. 
But as Chabanon and Rangamani showed previously, for a uniform distribution of proteins with no-flux boundary conditions on the membranes, minimal surfaces are admissible solutions for the membrane geometry \cite{chabanon2018,chabanon2019}.
In the case of closed geometries on vesicle, constant mean curvature surfaces are admissible solutions \cite{seifert1993curvature, gozdz1998composition,campelo2007model,ouyang1989bending}. 
In the case of interest here, a flat membrane is the admissible solution for the boundary conditions associated with $\tilde{z}$, and a proof of this result is given in Appendix~\ref{app:flatsurf}.%\todopr{David, after talking with Arijit, I got rid of the appendix. In previous papers we already explored the solution space for catenoids and helicoids and in this case, Arijit finds the flat plane (another minimal surface) to be the obvious solution. Let me know if you are okay with that.}

%However, further inspection shows that since the $\tilde{\sigma}$ boundary condition in both cases results in no-flux boundary conditions, $\tilde{\sigma}$ for the choice of boundary conditions for $\tilde{z}$ and $\tilde{\sigma}$, $\tilde{z}=0$ is the energy minimizing configuration (see Appendix for details). 


\subsection{Two-dimensional simulations\label{sec:2ddomain}}
\subsubsection{Numerical implementation}
\label{sec:numimp}
%\todopr[inline]{David, can you please check the following for accuracy?}

We solved the set of governing equations (\ref{eqn:nondgov_cont})--(\ref{eqn:nondgov_diff}) in two dimensions inside a square domain using a finite-difference technique that we outline here. %\todopr{Arijit, work that is already done should be written in past tense.}
Our numerical scheme is second order in space and first order in time. We note that time only appears explicitly in the advection-diffusion equation (\ref{eqn:nondgov_diff}) for the protein density: we solve it using a semi-implicit scheme wherein the linear diffusion term is treated implicitly while the nonlinear advective terms and curvature-induced transport terms are treated explicitly. The remaining governing equations are all elliptic in nature and can be recast as a series of Poisson problems as we explain. First, we note that the shape equation (\ref{eqn:nondgov_norn}) is biharmonic and can thus be recast into two nested Poisson problems providing the shape $\tilde{z}$ at a particular time step. To solve for the surface tension $\tilde{\lambda}$, we take the divergence of the tangential momentum balance (\ref{eqn:nondgov_tang}) and combine it with the continuity equation (\ref{eqn:nondgov_cont}) to obtain the Poisson equation 
\begin{equation}
  \tilde{\nabla}^2\tilde{\lambda}+f=0,  \label{eqn:poisson_pr}
\end{equation}
where
\begin{align}
\begin{split}
 f=&\,\,4\tilde{H}\tilde{\nabla}^2\tilde{w}-2\tilde{\boldsymbol{\nabla}}\tilde{\boldsymbol{\nabla}}\tilde{z}\boldsymbol{:}\tilde{\boldsymbol{\nabla}}\tilde{\boldsymbol{\nabla}} \tilde{w} -8\tilde{\boldsymbol{\nabla}} \tilde{H}\boldsymbol{\cdot}\tilde{\boldsymbol{\nabla}} \tilde{w} \\[2pt]
 &- \frac{{\vphantom{A}\smash{2 \hat{C} \hat{B}}}}{\hat{T}}\tilde{\boldsymbol{\nabla}} \tilde{\nabla}^2 \tilde{z}\boldsymbol{\cdot} \tilde{\boldsymbol{\nabla}} \tilde{\sigma}   
-\bigg[\frac{{\vphantom{A}\smash{2 \hat{C} \hat{B}}}}{\hat{T}} \tilde{\nabla}^2 \tilde{z}- \frac{2\hat{C}}{\hat{T}}\log \Big(\frac{\tilde{\sigma}}{\tilde{\sigma}_s} \Big) \bigg]\tilde{\nabla}^2 \tilde{\sigma} \\[-2pt]
&+ \frac{2\hat{C}}{\hat{T}}\tilde{\boldsymbol{\nabla}} (\log\tilde{\sigma}) \boldsymbol{\cdot} \tilde{\boldsymbol{\nabla}} \tilde{\sigma} +\hat{C} \hat{B}  \tilde{\nabla}^2 \tilde{\sigma}^2.    
\end{split}
\end{align}
Note that there is no natural boundary condition on $\tilde{\lambda}$ at the edges of the domain. To approximate an infinite membrane, we first estimate the 
tension along the four edges using the integral representation 
\begin{equation}
\label{eqn:int_rep_lam}
    \tilde{\lambda}(\tilde{\boldsymbol{r}})=1+\int_{\Omega}G(\tilde{\boldsymbol{r}}-\tilde{\boldsymbol{r}}_0)f(\tilde{\boldsymbol{r}}_0)\,\mathrm{d}A(\tilde{\boldsymbol{r}}_0),
\end{equation}
where $G(\boldsymbol{r})=-\log r/2\pi$ is the 2D Green's function for Poisson's equation in an infinite domain. The calculated tension along the edges is then used as the boundary condition for equation~(\ref{eqn:poisson_pr}), where the normal velocity component at the current time step $k$ is calculated as
\begin{equation}
    \tilde{w}^k=\frac{1}{Pe}\frac{\tilde{z}^k-\tilde{z}^{k-1}}{\Delta\tilde{t}}.
\end{equation}
With knowledge of the membrane tension, the tangential momentum balance (\ref{eqn:nondgov_tang}) then provides two modified Poisson problems for the in-plane velocity components. Note that the equations for $\tilde{z}$, $\tilde{\lambda}$ and $\tilde{\boldsymbol{v}}$ are nonlinearly coupled through the various forcing terms in their respective Poisson problems. To remedy this problem, we iterate their solution until every variable converges with a tolerance limit of $5\times 10^{-7}$ before proceeding to the next time step. All the results presented below were obtained on a spatial uniform grid of size $201\times 201$ and with a dimensionless time step of $\Delta\tilde{t}=10^{-4}$. We used Fortran 90 for compiling and running the algorithm. As we show in Appendix~\ref{app:numvalid}, the numerical method was successfully validated by comparison with a Stokes-Neumann formulation \cite{glowinski2005numerical}.

\subsubsection{2D simulation results}
\begin{figure}
    \centering
    \includegraphics[width=\textwidth]{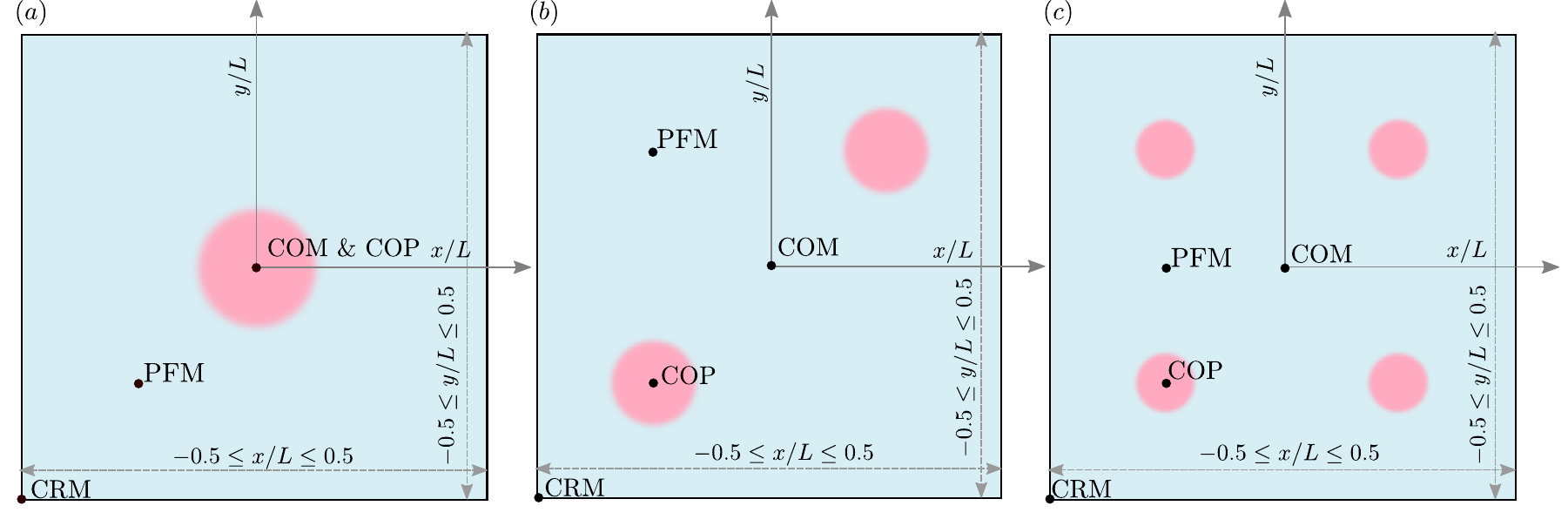}
    \caption{System set up and initial condition used in 2D simulations. All 2D simulations are performed for a linearized Monge patch. We simulate the dynamics for three different initial distributions of proteins as shown. The total area fraction of protein is same for the three cases, with proteins covering 10\% of the total area. (\textit{a}) Single patch of protein placed at the center $(0,0)$. (\textit{b}) Two patches of protein placed at diametrically opposite positions with center locations $(-0.25,-0.25)$ and $(0.25,0.25)$. (\textit{c}) Four patches of proteins placed at four diagonal positions: $(-0.25,-0.25)$, $(-0.25,0.25)$,$(0.25,-0.25)$ and $(0.25,0.25)$. The following abbreviations are used in subsequent figures to track the system behavior: COM: center of the membrane, COP: center of the patch, CRM: corner of the membrane, and PFM: protein-free membrane.}.
    \label{fig:schpatch}
\end{figure}
\begin{figure}
    \centering
    \includegraphics[width=\textwidth]{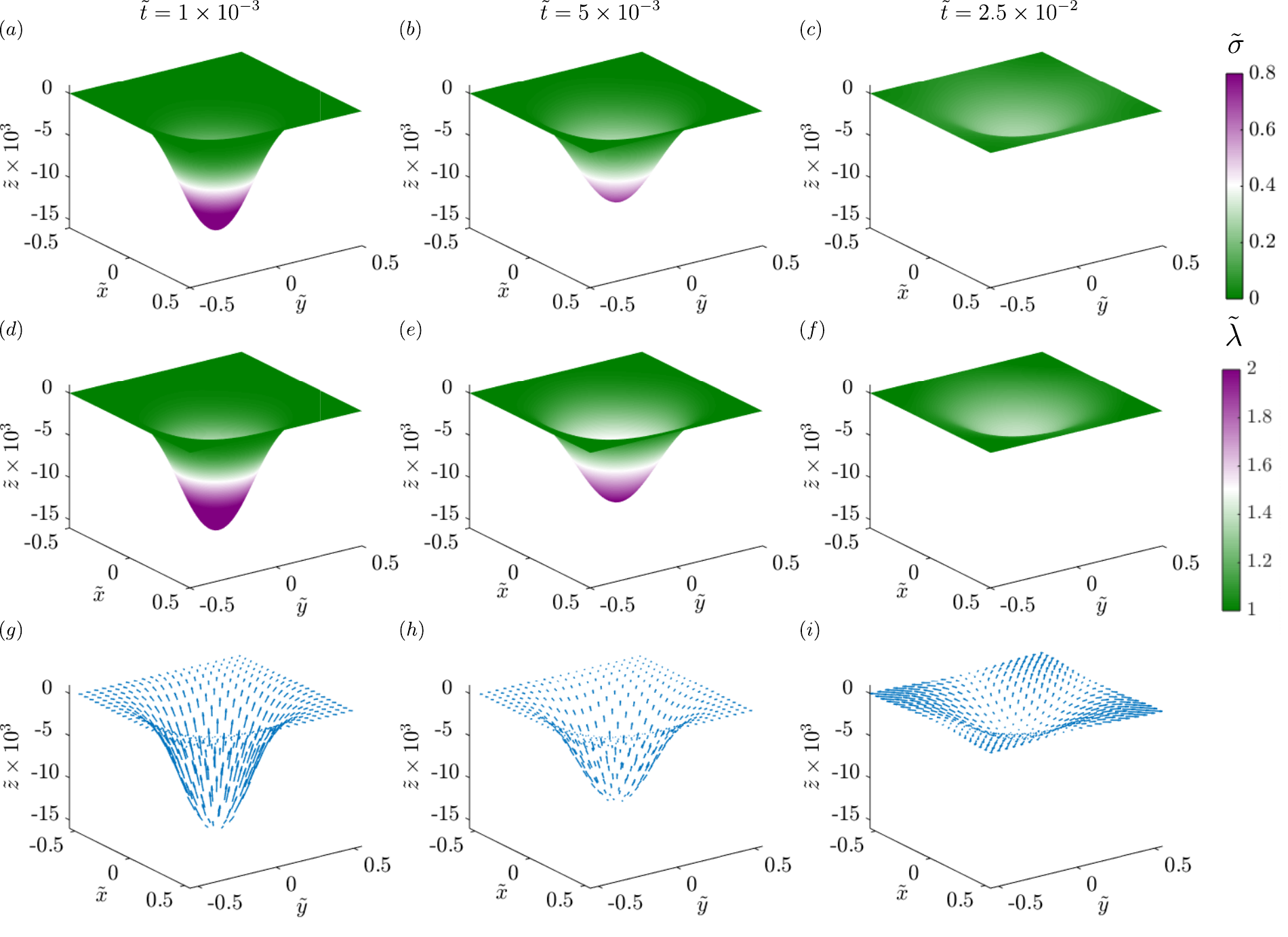}
    \caption{Dynamics of the evolution of membrane shape, protein distribution, membrane tension, and tangential velocity field for a single patch of protein at three different times. ($a$-$c$) Distributions of membrane protein density are shown at dimensionless times  $1\times10^{-3}$, $5\times10^{-3}$, and $2.5\times10^{-2}$. ($d$-$e$) Distributions of membrane tension at the same non-dimensional times. ($g$-$i$)  Tangential velocity fields shown at the same non-dimensional times. Arrows are scaled according to tangential velocity magnitude, with a maximum dimensionless velocity of $2.8\times10^{-3}$.}
    \label{fig:1csumm}
\end{figure}
%\todods[inline]{Format for labeling of figures and subfigures is figure 1(\textit{a}) (subfigures should be lower-case italics in parentheses). Regarding the velocity field in figure 4: it would be useful to somehow provide some information on the magnitude of the velocity. I seem to recall it's very small. }
%\todoam[inline]{Should I give the mmagnitute of the velocity in the figure or in the text? }
%\todopr[inline]{See my edit in the figure caption above. }
\begin{figure}
    \centering
    \includegraphics[width=\textwidth]{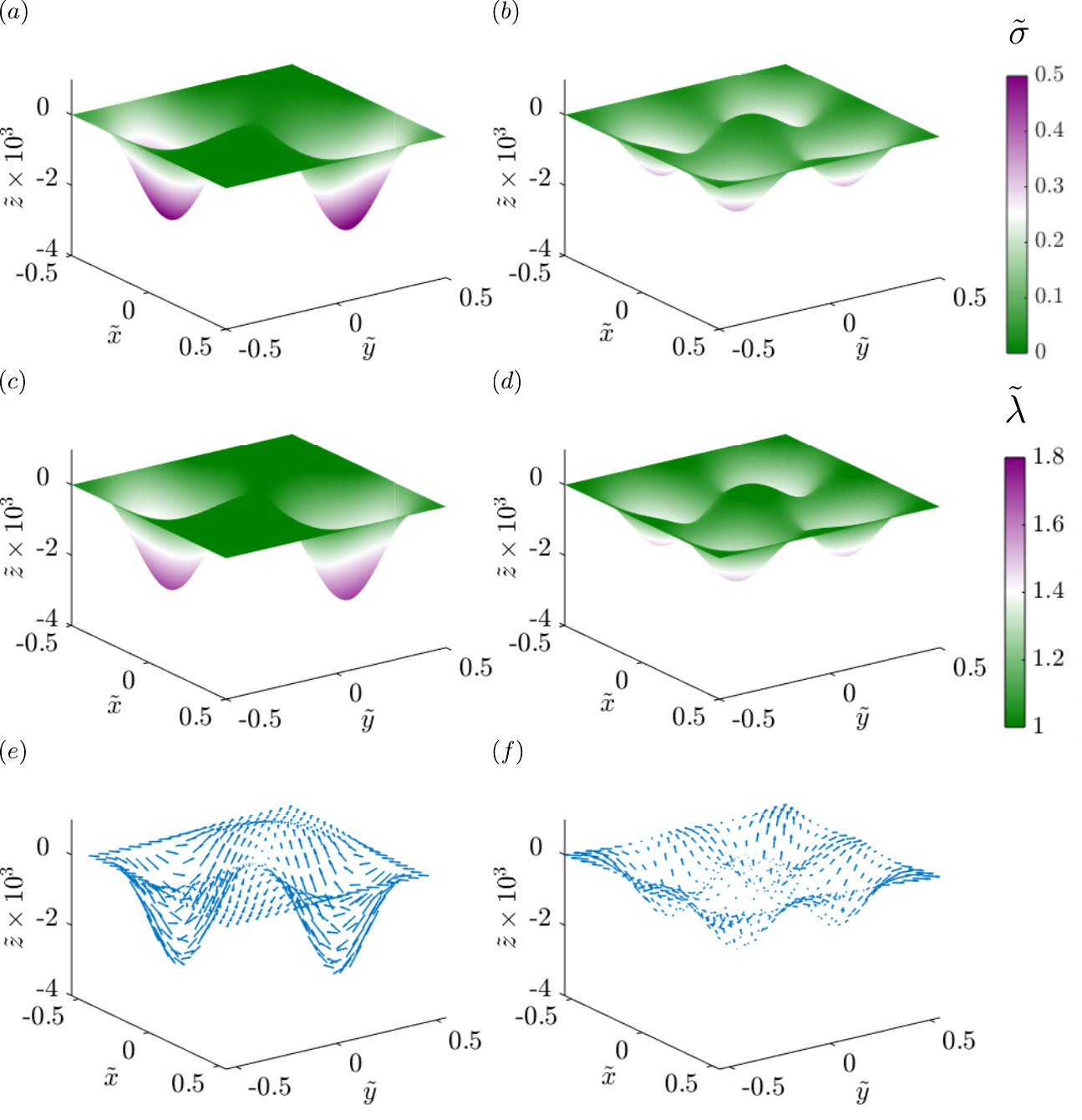}
    \caption{Dynamics of the evolution of membrane shape, protein distribution, membrane tension, and tangential velocity field for two and four patches of protein at dimensionless time $\tilde{t}=5 \times 10^{-3}$. The left column shows the distribution of protein density (\textit{a}), membrane tension (\textit{c}), and tangential velocity (\textit{e}) for two patches of protein. The right column shows the distribution of protein density (\textit{b}), membrane tension (\textit{d}), and tangential velocity (\textit{f}) for four patches of protein. The magnitude of maximum dimensionless tangential velocity is $2.2\times10^{-2}$ in the case of two patches, and $4.1\times10^{-3}$ in the case of four patches.}
    \label{fig:2c4cint}
\end{figure}
\begin{figure}
    \centering
    \includegraphics[width=\textwidth]{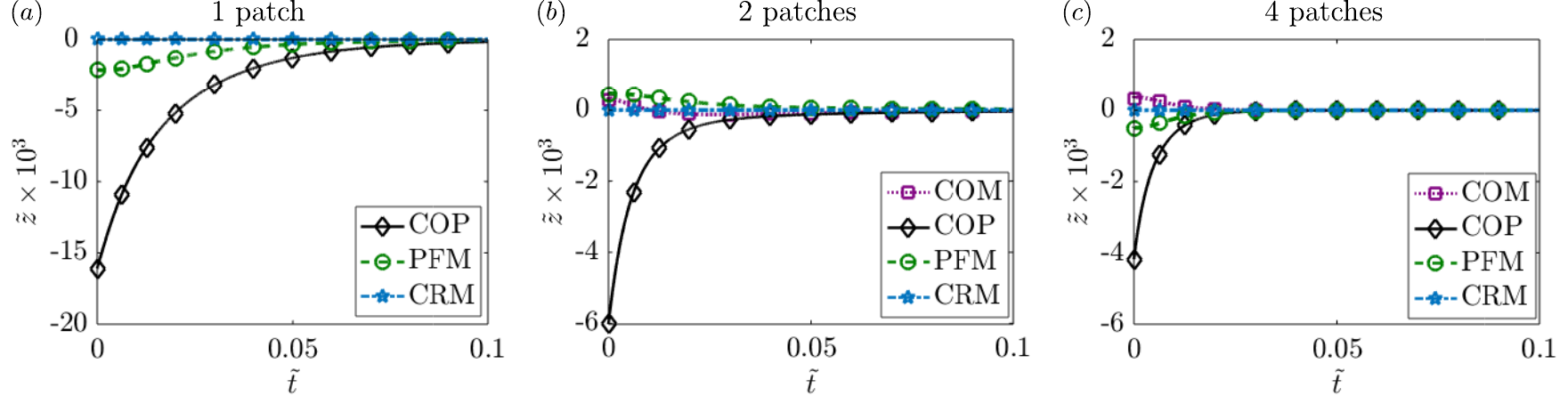}
    \caption{Temporal evolution of the membrane deflection at the various locations defined in figure~\ref{fig:schpatch} for a single patch of protein (\textit{a}), two patches (\textit{b}), and four patches (\textit{c}).}
    \label{fig:zlocvar}
\end{figure}
\begin{figure}
    \centering
    \includegraphics[width=\textwidth]{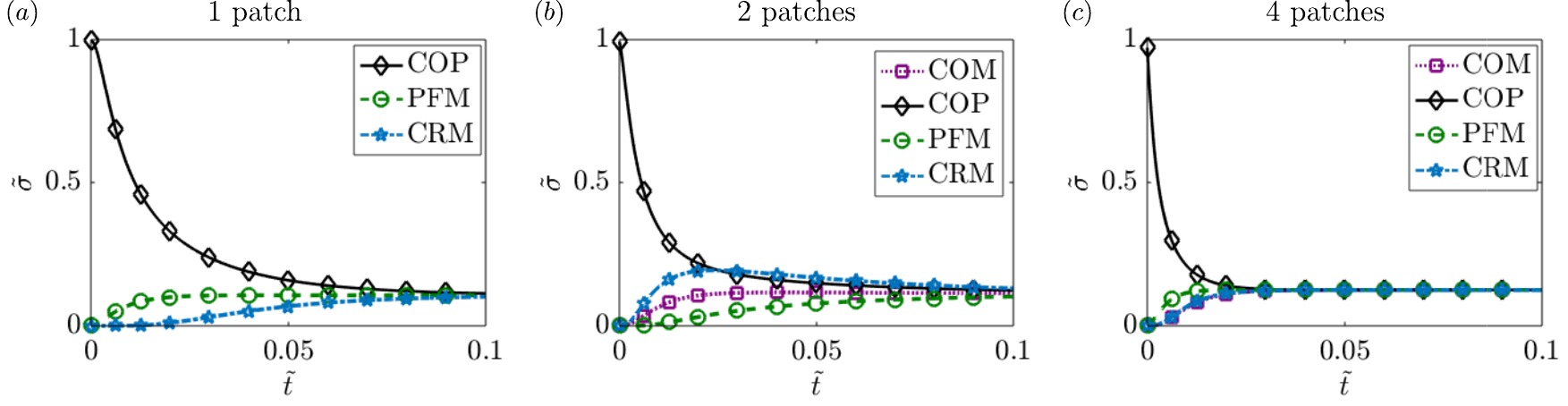}
    \caption{Temporal evolution of the protein density at the locations defined in figure~\ref{fig:schpatch} for a single patch of protein (\textit{a}), two patches  (\textit{b}), and four patches (\textit{c}).}
    \label{fig:siglocvar}
\end{figure}
\begin{figure}
    \centering
    \includegraphics[width=\textwidth]{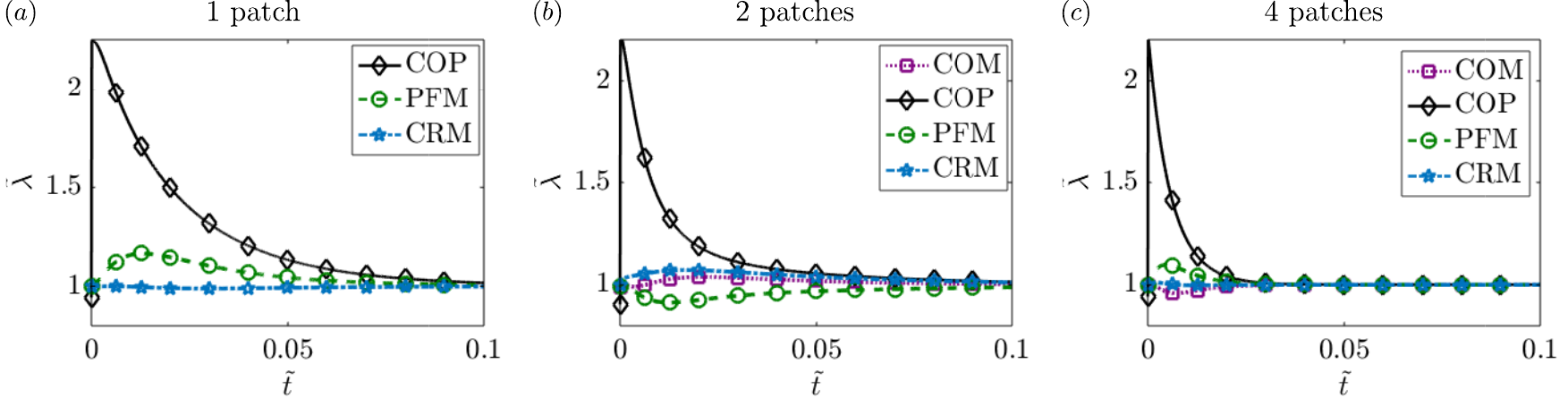}
    \caption{Temporal evolution of the membrane tension at the locations defined in figure~\ref{fig:schpatch} for a single patch of protein (\textit{a}), two patches  (\textit{b}), and four patches (\textit{c}).}
    \label{fig:lamlocvar}
\end{figure}

Using the numerical scheme described above, we solved the linearized two-dimensional governing equations (\ref{eqn:nondgov_cont})--(\ref{eqn:nondgov_diff}) for different initial conditions. %\todopr{Check equation cross reference}
In all cases, the boundary conditions for the membrane shape were set to $\tilde{z}=0$ and $\boldsymbol{\nu}\boldsymbol{\cdot}\boldsymbol{\nabla} \tilde{z} =0$ where $\boldsymbol{\nu}$ is the in-plane normal to the  edge of the domain, and no-flux boundary conditions were enforced on the protein distribution. We considered three different initial conditions for the protein density as depicted in figure~\ref{fig:schpatch}, namely: a single circular patch at the center of the domain (figure~\ref{fig:schpatch}($a$)), two indentical patches placed at diametrically opposite ends of the domain (figure~\ref{fig:schpatch}($b$)), and four patches centered in each quadrant of the domain (figure~\ref{fig:schpatch}($c$)). 
The total mass of protein is the same in all three cases, only the initial spatial distribution is different. For the velocity and tension, we maintain open boundary conditions as noted in \S\,\ref{sec:numimp}. 

We tracked the dynamics of the membrane shape, protein distribution, membrane tension, and velocity for a single patch of proteins corresponding to figure~\ref{fig:schpatch}($a$) in figure~\ref{fig:1csumm}. 
The initial membrane configuration is bent to accommodate the initial distribution of proteins (figure~\ref{fig:1csumm}($a$)), and the membrane tension for this initial distribution is heterogeneous as seen in figure~\ref{fig:1csumm}($d$), consistent with our previous results \cite{Rangamani14,Hassinger17}. Over time, the proteins diffuse from the center of the patch across the membrane, tending towards a homogeneous distribution (figure~\ref{fig:1csumm}($b,c$)), and this process is accompanied by a reduction in the membrane deflection. 
The homogenization of proteins results in homogenization of the membrane tension, which approaches its value at infinity (figure~\ref{fig:1csumm}($e,f$)). 
The tangential velocity is directed outward (figure~\ref{fig:1csumm}($g,h,i$)), and the dimensionless magnitude of the maximum velocity in figure~\ref{fig:1csumm}(\textit{g}) is $4.8 \times 10^{-3}$.
Expectedly, the magnitude of this velocity decreases with time as seen in figure~\ref{fig:1csumm}($h,i$).

When the proteins are distributed in two and four patches as shown in figure~\ref{fig:schpatch}($b,c$), we found that the overall behavior of the system was quite similar to a single patch with some changes to the dynamics.
First, because each patch had a lower density of proteins (half or quarter), the initial deformation was smaller and the protein distribution homogenized faster than in the case of a single patch (figure~\ref{fig:2c4cint}($a,b$)).
Similarly, the typical magnitude of membrane tension variations (figure~\ref{fig:2c4cint}($c,d$)) and of the tangential velocity field (figure~\ref{fig:2c4cint}$(e,f)$) was also smaller to begin with and the system attained the homogeneous distribution rapidly. 
%\Cref{fig:1csumm} shows the distribution of membrane parameters projected on the plane of the membrane at three different times for the initial condition of a single patch of protein (\ref{fig:schpatch}(\textit{a})). It shows as the time approaches protein diffuses along the plane of the membrane, membrane deformation decreases accordingly. This is consistent with the result of 1-D string case. We also find that the membrane tension approaches to the tension at the infinity. Direction of the tangential velocity is outward to the domain and magnitude of velocity diminishes with time. \Cref{fig:2c4cint} distribution of membrane density, membrane tension and velocity vectors on the plane of the deformed membrane at the dimensionless time of $0.005$ corresponding to the initial conditions of two and four patches of protein (\Cref{fig:schpatch}(\textit{b}),(\textit{b})). We find that density of protein and membrane tension is more homogeneous at the case of four patches compared to two patches. We further notice that velocity magnitude is higher at the patch free end for the two patches of protein.   

To compare the effects of one, two, and four patches directly, we plotted the membrane deformation (figure~\ref{fig:zlocvar}),  membrane protein distribution (figure~\ref{fig:siglocvar}), and membrane tension distribution (figure~\ref{fig:lamlocvar}) at different locations for each case. %\todopr{For figures 6-8, please consider adding a panel title one patch, 2 patches, and 4 patches on top of each panel.}
The initial deformation is different for the different cases because of the differences in the local density of proteins. 
For a single patch, we observed that  the maximum deformation occurs at the center of the patch (COP) and it takes a longer time for this deformation to go to zero in the case of a single patch compared to multiple patches (compare figure~\ref{fig:zlocvar}($a$) to figure~\ref{fig:zlocvar}($b,c$)). 
In the case of two and four patches, we also observe a small positive deformation at the center of the membrane (COM) and in the protein-free membrane (PFM). 
This can be explained by the fact that the continuity conditions of the surface will result in a small but upward displacement in protein-free regions in response to the large downward displacement in the regions where the proteins are initially present. 

Comparing the protein dynamics for one, two, and four patches, we observed that increasing the number of patches decreases the time it takes for the protein distribution to homogenize across the membrane domain (figure~\ref{fig:siglocvar}).
Thus, although membrane bending and protein distribution are coupled, the distribution of multiple patches weakens the coupling and promotes rapid homogenization of the membrane proteins. 
While the steady state protein distribution is the same in all cases, the dynamics with which the protein-free regions show an appreciable increase in proteins also depends on the initial distribution of proteins. 
For example, in the case of a single patch, CRM takes much longer to reach steady state compared to the case of two or four patches. 
\Cref{fig:siglocvar} shows that for the initial conditions of a single patch and of four patches of protein, $\tilde{\sigma}$ approaches the uniform protein density monotonically. 
But for the case of two patches, the time evolution of the protein density is not monotonic, and  
we found the density of protein at the corner of the membrane (CRM) exceeds the density at the center of the patch (COP) for a brief time interval.
%This is because the coupling of lipid flow and protein diffusion results in convective transport of proteins towards the corner of the membrane.
We investigated this phenomenon further and studied the dependence on intra-membrane flow by varying the P\'eclet number in \S
\,\ref{sec:Pe_dependene}.

Similar dynamics are observed for the membrane tension as well.
\Cref{fig:lamlocvar} shows that the membrane tension takes a larger time to reach its steady value for the case of one or two patches when compared to four patches (compare figures~\ref{fig:lamlocvar}($a,b,c$)).
The initial rise in the membrane tension corresponds to the inviscid response of the membrane to the curvature-inducing protein distribution, while from the next time step onwards tension changes primarily due to viscous effects.

\subsubsection{Effect of fluid advection on coupled membrane--protein dynamics}

\label{sec:Pe_dependene}
\begin{figure}
    \centering
    \includegraphics[width=\textwidth]{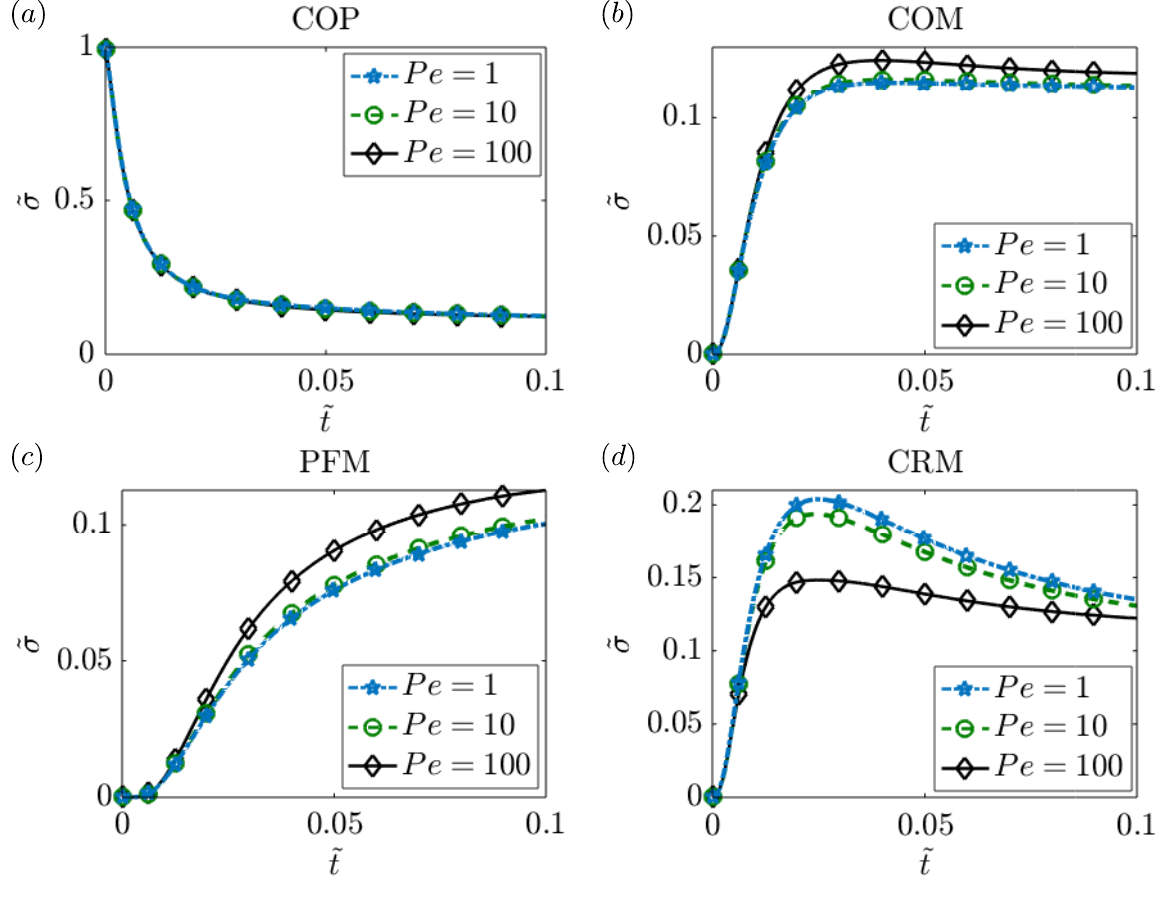}
    \caption{Temporal evolution of local protein density for different values of the P\'eclet number in the case of two patches of protein. The protein density is measured at the four locations defined in figure~\ref{fig:schpatch}: COM (\textit{a}), COP (\textit{b}), PFM (\textit{c}) and CRM (\textit{d}).}
    \label{fig:pec}
\end{figure}
\begin{figure}
    \centering
    \includegraphics[scale=.6]{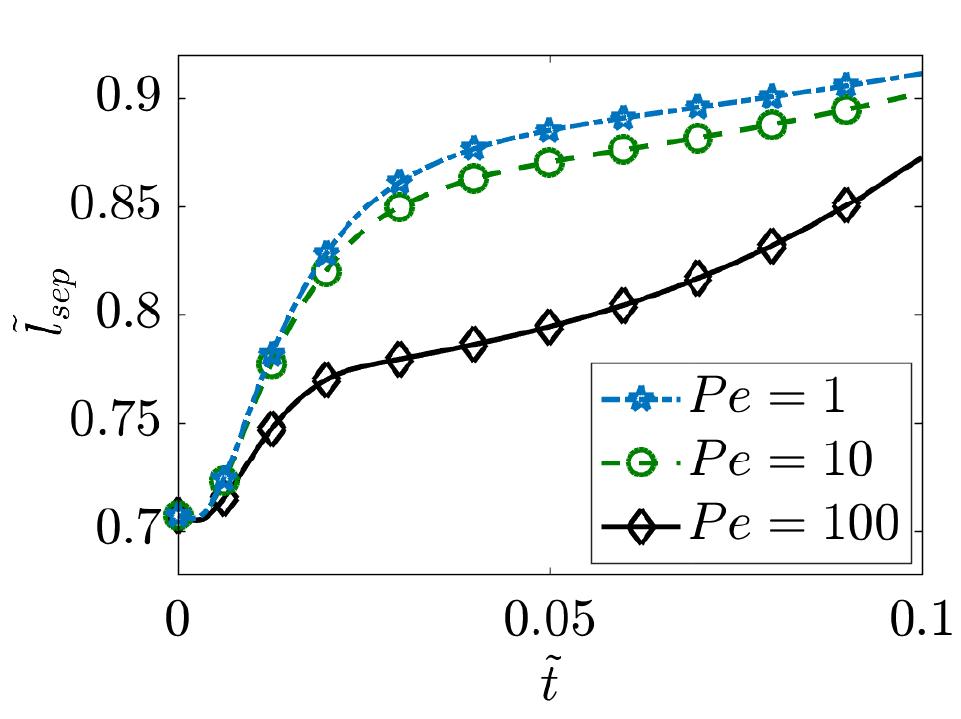}
    \caption{Evolution of the separation distance between the centroids of the protein patches in the case of two patches and for three different values of the P\'eclet number.}
    \label{fig:vel2c}
\end{figure}
Next, we investigated the effect of fluid advection in the case of two patches by varying the P\'eclet number $Pe$ in figure~\ref{fig:pec}.
We observed that at the center of the patch (figure~\ref{fig:pec}($a$)) there was no observable effect on the temporal evolution of the protein density. 
However, at the center of the membrane (figure~\ref{fig:pec}($b$)) and in the protein-free membrane (figure~\ref{fig:pec}($c$)), we observed that increasing P\'eclet number had a small effect on the dynamics of the protein density, particularly at long times.

The effect of increasing the P\'eclet number was most dramatic at the corner of the membrane (figure~\ref{fig:pec}($d$)), where the initial rise in the protein density was found to be similar for all three values of \textit{Pe}, but the increase resulted in a higher value for lower {Pe}. Eventually, $\tilde{\sigma}$ at the corner decreases towards the mean value of $\tilde{\sigma}$ over time. 
Thus, the coupling between lipid flow and protein diffusion seems to have a larger impact on transport in the regions that are initially protein-free. 

%We noticed that as we increased the P\'eclet number, the difference between the density of protein at the corner (CRM) and at the center of the initial location of the patch (COP) increases. 
%This shows that the flow of lipid causes increased advection transport of protein towards the corner more with an increase in P\'eclet number. 

To further investigate the role of convective transport, we tracked the separation distance $\tilde{l}_{sep}$ between the centers of mass of two effective patches ($l_{sep}$) as a function of time in figure~\ref{fig:vel2c}. The center of mass of a patch is formally defined as
\begin{equation}
\label{eqn:centroid1}
    \tilde{\boldsymbol{r}}_c=\frac{\displaystyle\int_\Omega \tilde{\boldsymbol{r}}(\tilde{\sigma}-\tilde{\sigma}_m) \mathcal{H}(\tilde{\sigma}-\tilde{\sigma}_m) \,\mathrm{d}{a}}{\displaystyle \int_\Omega (\tilde{\sigma}-\tilde{\sigma}_m)\mathcal{H}(\tilde{\sigma}-\tilde{\sigma}_m)\,\mathrm{d}{a}}  \quad \text{with,} \quad \tilde{\sigma}_m=\int_\Omega \tilde{\sigma}\,\mathrm{d}{a},
\end{equation}
where the effective extent of the patch is defined using the Heaviside function $\mathcal{H}$ as the area where protein density exceeds its mean value.  %\todopr{Arijit, say why you need this.}
%The distance between two effective patches, $l_{sep}$ is then given by \todopr{Arijit, finish this}.
We observed that $\tilde{l}_{sep}$ increases with time and decreases with increasing P\'eclet number (figure~\ref{fig:vel2c}).
This can be explained from the velocity profile for two patches in figure~\ref{fig:2c4cint}(\textit{e}). The direction of the velocity is towards the center of the membrane in the area where the patch is located. Therefore, the advective transport due to the lipid tends to weaken the separation otherwise caused by diffusion. Since the effect of flow increases with increasing $Pe$, the separation of the two patches slows down for higher values P\'eclet number as shown in figure~\ref{fig:vel2c}. This also explains the decrease of density of the protein at the corner of the membrane (CRM) and the increase of the protein density at the initial protein-free area (PFM) and center of the membrane (COM) with higher value of \textit{Pe} as found in figure~\ref{fig:pec}. 

We can further understand the dynamics of the separation distance between the two patches by considering the diffusion of a protein patch in one triangular half domain of lipid.
This triangle is bounded by two of the domain boundaries and by the diagonal of the square domain that passes in between the two patches. 
The diagonal line is also a line of symmetry, and thus behaves as an effective no-flux boundary for the triangular half of the domain. 
Therefore, each triangular half-domain is subject to the no-flux condition on its three sides. 
In this half domain, the semicircular half patch of protein facing the corner of the membrane (CRM) diffuses to a smaller area compared to the other semicircle that faces the center of the membrane (COM). 
This results in an effectively larger protein gradient towards CRM.
Therefore, the protein density shifts towards the protein-free corner and results in an effective shift of the patches towards the corners of the membrane.

\section{Axisymmetric membranes} 
\label{sec:axisymmetry}
In the previous section, we focused on small deformations from a flat plane. However, membranes are known to undergo large deformations including bud-like shapes in the presence of proteins \cite{Hassinger17, haucke2018membrane, dannhauser2012reconstitution}. % \todopr{need references from other groups too, we aren't the only ones to model buds}.
Here, we illustrate the interaction between membrane bending and protein diffusion for membrane buds. For these simulations, we assume that the membrane is rotationally symmetric and recast the governing equations of \S \ref{sec:gov_eqn} in an axisymmetic framework.

\subsection{Governing equations in axisymmetric coordinates}

We represent tangential velocity vector in polar coordinates as
\begin{equation}
    \boldsymbol{v}=v^s \boldsymbol{e}_s+ v^{\theta} \boldsymbol{e}_{\theta}.
\end{equation}
For an axisymmetric geometry as depicted in figure~\ref{fig:axi_zoom_out}($a$), we assume $\partial() /\partial \theta=0$ and $v^{\theta}=0$ \cite{Arroyo09}.
Thus, we parametrize the geometry as
\begin{equation}
    \boldsymbol{r}(s,\theta, t)=r(s,t)\boldsymbol{e}_r+z(s,t)\boldsymbol{k},
\end{equation}
where the unit vectors $(\boldsymbol{e}_r,\boldsymbol{e}_{\theta},\boldsymbol{k})$ are a set of orthonormal basis vectors and $s$ is the arclength measured from the axis of symmetry.
The tangent and normal vectors are given by
\begin{align}
\boldsymbol{a}_s= \cos \psi \boldsymbol{e}_{r}+ \sin \psi \boldsymbol{e}_{\theta} , \quad \boldsymbol{a}_{\theta}=r\boldsymbol{e}_{\theta}, \quad \boldsymbol{n}=-\sin \psi \boldsymbol{e}_{r}+ \cos \psi \boldsymbol{e}_{\theta},
\end{align}
where $\psi$ is the angle made by the tangent $\boldsymbol{a}_s$ with the radial unit vector $\boldsymbol{e}_r$. The corresponding surface metric ($a_{\alpha \beta}$) and curvature metric ($b_{\alpha \beta}$) are
\begin{equation}
   a_{\alpha \beta}=
  \left[ {\begin{array}{cc}
   1 & 0 \\
   0 & r^2 \\
  \end{array} } \right], \quad
  b_{\alpha \beta}=
  \left[ {\begin{array}{cc}
   \psi_{,s} & 0 \\
   0 & r \sin \psi \\
  \end{array} } \right].
\end{equation}

Using these expressions, the incompressibility constraint becomes
\begin{equation}
\label{eqn:axicontnd}
\frac{1}{r}\frac{\partial (r v^s)}{\partial s}=2wH.
\end{equation}
The governing equation for surface pressure is
\begin{equation}
\begin{split}
\frac{\partial \lambda}{\partial s}-4\nu w\frac{\partial H}{\partial s}+2\nu \bigg(2 \frac{\partial wH}{\partial s}+Kv_s  &-\frac{\partial w}{\partial s} \frac{\partial \psi}{\partial s} \bigg)
 \\ &= -\frac{\partial \sigma}{\partial s} \bigg[ k_BT\log\Big(\frac{\sigma}{\sigma_s}\Big) -2k \ell (H-\ell \sigma) \bigg], 
%2A_1 (A_1 \sigma +A_0)&\boldsymbol{\nabla} \sigma, 
\end{split}
\end{equation}
%\todods{Provide expression of $\boldsymbol{b}$ for Monge}
%\todoam{I have used the expression of $\boldsymbol{b}$ here directly, will this work?}
%\todopr{Arijit, give the expression of both $\boldsymbol{a}$ and $\boldsymbol{b}$ earlier after you define the tangents and normals. This should be after equation 3.2}
and the shape equation expressing the normal momentum balance is given by
\begin{align}
\begin{split}
   k\frac{1}{r} \frac{\partial}{\partial s} \left(r \frac{\partial  (H-\ell\sigma)}{\partial s} \right)&+2k(H-\ell\sigma)(2H^{2}-K) \\
&    -2H\bigg[k_BT\sigma\Big(\log\Big(\frac{\sigma}{\sigma_s}\Big)-1\Big)+k(H-\ell\sigma)^{2}\bigg] \\
& - 2\nu \bigg[  \frac{\partial \psi }{\partial s}  \frac{\partial v^s }{\partial s} + \frac{\sin \psi \cos \psi v^s}{r^2}- w(4H^{2}-2K)\bigg] =p+2\lambda H.  
\end{split}\label{eqn:axishapend}
\end{align}
The transport equation for the protein density simplifies to
\begin{equation}
\begin{split}
\frac{\partial {\sigma}}{\partial t}+ \frac{\partial (\sigma v^s) }{\partial s}=D  \frac{1}{r}\frac{\partial}{\partial s}  \left(r\frac{\partial \sigma}{\partial s} \right)&-2k\ell \bigg[\sigma \frac{1}{r}\frac{\partial}{\partial s}  \left(r\frac{\partial H}{\partial s} \right) +\left(\frac{\partial H}{\partial s} \right)^2\bigg] \\ 
&+2k\ell^2\bigg[\sigma \frac{1}{r}\frac{\partial}{\partial s}  \left(r\frac{\partial \sigma}{\partial s} \right) +\left(\frac{\partial \sigma}{\partial s} \right)^2\bigg].
\end{split}
\end{equation}
Finally, the kinematic relation for the normal velocity is given by
\begin{equation}
\label{eqn:axiwnd}
w=\boldsymbol{n}\boldsymbol{\cdot}\frac{\partial \boldsymbol{r}}{\partial t}.
\end{equation}

\subsection{Non-dimensionalization}

We non-dimensionalize the system of equations using a reference length scale $L$ (which we assumed to be 20 nm) such that the radius of the domain is $20L$, with all other scales remaining the same as in \S \ref{sec:nond_Monge}. 
The dimensionless governing equations are:
\begin{align}
\label{eqn:nondgov_cont_axi}
& \frac{1}{\tilde{r}} \frac{\partial (\tilde{r}\tilde{v}^s)}{\partial \tilde{s}} =2\tilde{w}\tilde{H}, \\[4pt]
\begin{split}
\label{eqn:nondgov_tang_axi}
\frac{\partial \tilde{\lambda}}{\partial \tilde{s}}-4 \tilde{w}\frac{\partial \tilde{H}}{\partial \tilde{s}}+& 2 \bigg(2 \frac{\partial \tilde{w} \tilde{H}}{\partial \tilde{s}}+K\tilde{v}_s  -\frac{\partial \tilde{w}}{\partial \tilde{s}} \frac{\partial \psi}{\partial \tilde{s}} \bigg)
= \\ 
&-\frac{\partial \tilde{\sigma}}{\partial \tilde{s}} \bigg[  \frac{2 \hat{C}}{\hat{T}}\log\Big(\frac{\tilde{\sigma}}{\tilde{\sigma}_s}\Big) - \frac{4 \hat{C} \hat{B}}{\hat{T}} (\tilde{H}-\hat{C} \hat{B} \tilde{\sigma}) \bigg],
\end{split} \\[4pt]
\begin{split}
\label{eqn:nondgov_norn_axi}
\frac{1}{\tilde{r}} \frac{\partial}{\partial \tilde{s}} \bigg(\tilde{r} \frac{\partial  ( \tilde{H}-\hat{C} \hat{B}\tilde{\sigma})}{\partial \tilde{s}} \bigg)& + 2(\tilde{H}-\hat{C}\hat{B} \tilde{\sigma})( 2\tilde{H}^{2}- K)  \\
 -2\tilde{H}&\bigg[\hat{C}\tilde{\sigma}\Big(\log\Big(\frac{\tilde{\sigma}}{\tilde{\sigma}_s}\Big)-1\Big)+(\tilde{H}-\hat{C} \hat{B} \tilde{\sigma})^{2}\bigg]   \\
 -  \hat{T} \bigg[ &\frac{\partial \psi}{\partial \tilde{s}} \frac{\partial v^s}{\partial \tilde{s}} + \frac{\sin \psi \cos \psi \tilde{v}^s}{\tilde{r}^2}- \tilde{w}(4\tilde{H}^{2}-2\tilde{K})\bigg] =\frac{\hat{P}}{2}+\hat{T} \tilde{\lambda} \tilde{H},
 \end{split} \\[7pt]
 \label{eqn:nondgov_diff_axi}
 \begin{split}
  \frac{\partial \tilde{\sigma}}{\partial \tilde{t}}+ Pe \frac{\partial (v^{\alpha} \tilde{\sigma})}{\partial \tilde{s}} & = \frac{1}{\tilde{r}}\frac{\partial }{\partial \tilde{s}}\left(\tilde{r} \frac{\partial \tilde{\sigma}}{\partial \tilde{s}}\right)\\
    & -2 \hat{B} \bigg[\frac{\partial (\tilde{H}-\hat{C} \hat{B} \tilde{\sigma})}{\partial \tilde{s}} \frac{\partial \tilde{\sigma}}{\partial \tilde{s}}+\tilde{\sigma} \frac{1}{\tilde{r}} \frac{\partial }{\partial \tilde{s}}\bigg( \tilde{r}\frac{\partial (\tilde{H}-\hat{C} \hat{B} \tilde{\sigma})}{\partial \tilde{s}}\bigg)\bigg],
 \end{split} \\[7pt]
    &\tilde{w}=\frac{1}{Pe}\boldsymbol{n}\boldsymbol{\cdot}\frac{\partial\tilde{\boldsymbol{r}}}{\partial \tilde{t}}. \label{eqn:kin_nondw_axi}
\end{align}
%\todopr{are there any new terms that need to be defined? or everything the same as before?}
%\todoam{Everything is same as before}
%The dimensionless chemical potential $C_h={A_0}/{A_1 \sigma_0}$ represents the ratio between the reference chemical potential of the membrane and the protein-induced chemical potential. 
  %leaving four dimensionless groups in the problem. 
  
%\subsection{Numerical simulations}
%\todopr{David, can you please take a close look?}
\subsection{Numerical implementation}
The system of dimensionless governing equations (\ref{eqn:nondgov_tang_axi})--(\ref{eqn:kin_nondw_axi}) was solved using finite difference methods using the arclength parametrization \cite{Rangamani20}.
%\todopr{You can reference Ali's paper here}
In these simulations, the total area of the domain was kept constant. 
This was achieved by first dividing the initial arclength into $N-1$ discrete elements, which gives $N$ grid points on which the equations were solved.
We then calculated the area of each of the $N-1$ discrete area elements.
At each time step, the local radius and arclength is back-calculated by keeping the area of these discrete elements constant in the following way: %\needref \todoam{There is no reference for this, instead of that I have added the formulation}
\begin{equation}
    r_{i+1}^2=r_i^2+2 \mathrm{d}A_i \cos \psi /\pi,
\end{equation}
and
\begin{equation}
   s_{i+1}=s_i+\mathrm{d}s_i=s_i+\frac{r_{i+1}-r_i}{\cos \psi},
\end{equation}
with
\begin{equation}
    r_1=s_1=0.
\end{equation}
To solve the set of governing equations (\ref{eqn:nondgov_tang_axi})--(\ref{eqn:kin_nondw_axi}), we first obtained the membrane tension $\tilde{\lambda}$ by integrating equation (\ref{eqn:nondgov_tang_axi}) backward starting from the edge of the membrane where the boundary condition $\tilde{\lambda}=1$ is enforced. 
We then consider the shape equation (\ref{eqn:nondgov_norn_axi}) where the first term can be written as $(1/\tilde{r})(\partial \tilde{L}/\partial \tilde{s})$ in terms of the normal bending stress \cite{Steigmann03}
\begin{equation}
\label{eqn:nond_psi}
    \frac{\tilde{L}}{\tilde{r}}=\frac{\partial }{\partial \tilde{s}} \left[\frac{1}{2}\left( \psi_s+\frac{\sin \psi }{\tilde{r}} \right) -\hat{C} \hat{B} \tilde{\sigma} \right].
\end{equation} 
The modified shape equation is solved for $\tilde{L}$ with boundary condition $\tilde{L}=0$ at the center of the domain
corresponding to the case where there is no pulling force acting on the center of the membrane. When doing do, other shape-dependent terms in the shape equation are treated explicitly, and iterations are performed until convergence.
Equation (\ref{eqn:nond_psi}) is then integrated for $\psi$ at every point with the boundary condition $\psi=0$ at the center of the membrane and at the boundary. 
Having determined $\psi$, the radial $\tilde{r}$ and the vertical $\tilde{z}$ position of the membrane are calculated. 
The continuity equation (\ref{eqn:nondgov_cont_axi}) is then integrated to obtain the value of tangential velocity $\tilde{v}^s$. 
Finally, the diffusion equation (\ref{eqn:nondgov_diff_axi}) is marched in time to update the protein distribution $\tilde{\sigma}$ across the membrane as described in \S \ref{sec:numimp}.

\subsection{Numerical results}
\begin{figure}
    \centering
    \includegraphics[width=\textwidth]{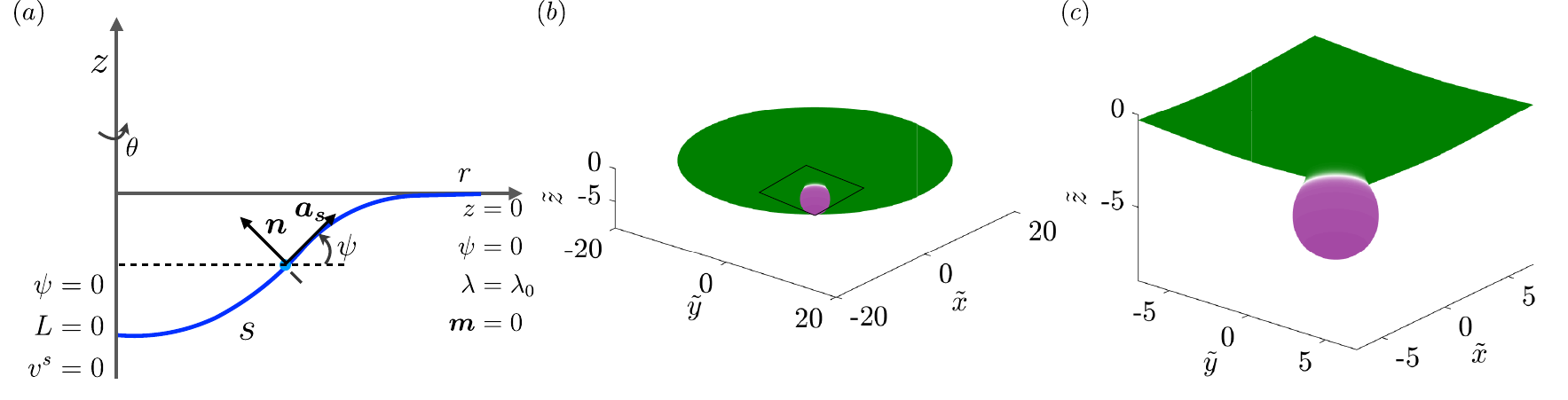}
    \caption{Parametrization of an axisymmetric membrane and bud-shaped initial condition: (\textit{a}) Parametrization and boundary conditions  for an axisymmetric membrane. (\textit{b}) Solution domain showing the initial condition, where a circular patch of curvature-inducing protein (shown in purple) induces a bud-shaped deformation.  (\textit{c}) A magnified view of the domain shown in (\textit{a}).}
    \label{fig:axi_zoom_out}
\end{figure}

We solved the dimensionless governing equations (\ref{eqn:nondgov_cont_axi})--(\ref{eqn:kin_nondw_axi}) for the
 solution domain and boundary conditions shown in figure~\ref{fig:axi_zoom_out}(\textit{a}).
The domain is initialized with a protein distribution as shown in figure~\ref{fig:axi_zoom_out}(\textit{b}) such that the initial shape of the membrane is a bud, similar to those observed in membrane fission and fusion processes \cite{kozlovsky2003fission, jahn2003fusion}. 
This initial shape is obtained by solving equations (\ref{eqn:axicontnd})--(\ref{eqn:axiwnd}) %\todopr{cross ref equations} 
for an inviscid membrane.
The black square highlights the curved bud region shown as a zoomed-in image in figure~\ref{fig:axi_zoom_out}(\textit{c}), and all simulation results are shown in this zoomed-in region.

We make no simplifying assumptions about linear deformations or curvature regimes in the axisymmetric case, which allows us to explore the nonlinear coupling between membrane curvature, protein diffusion, and lipid flow in full detail. 
We conducted the following simulations to map this relationship: (a) Diffusion of proteins on curved surfaces with no coupling between protein distribution and spontaneous curvature, so as to study the effect of surface curvature on protein diffusion, and (b) coupled diffusion of proteins and induction of spontaneous curvature.

\begin{figure}
    \centering
    \includegraphics[width=\textwidth]{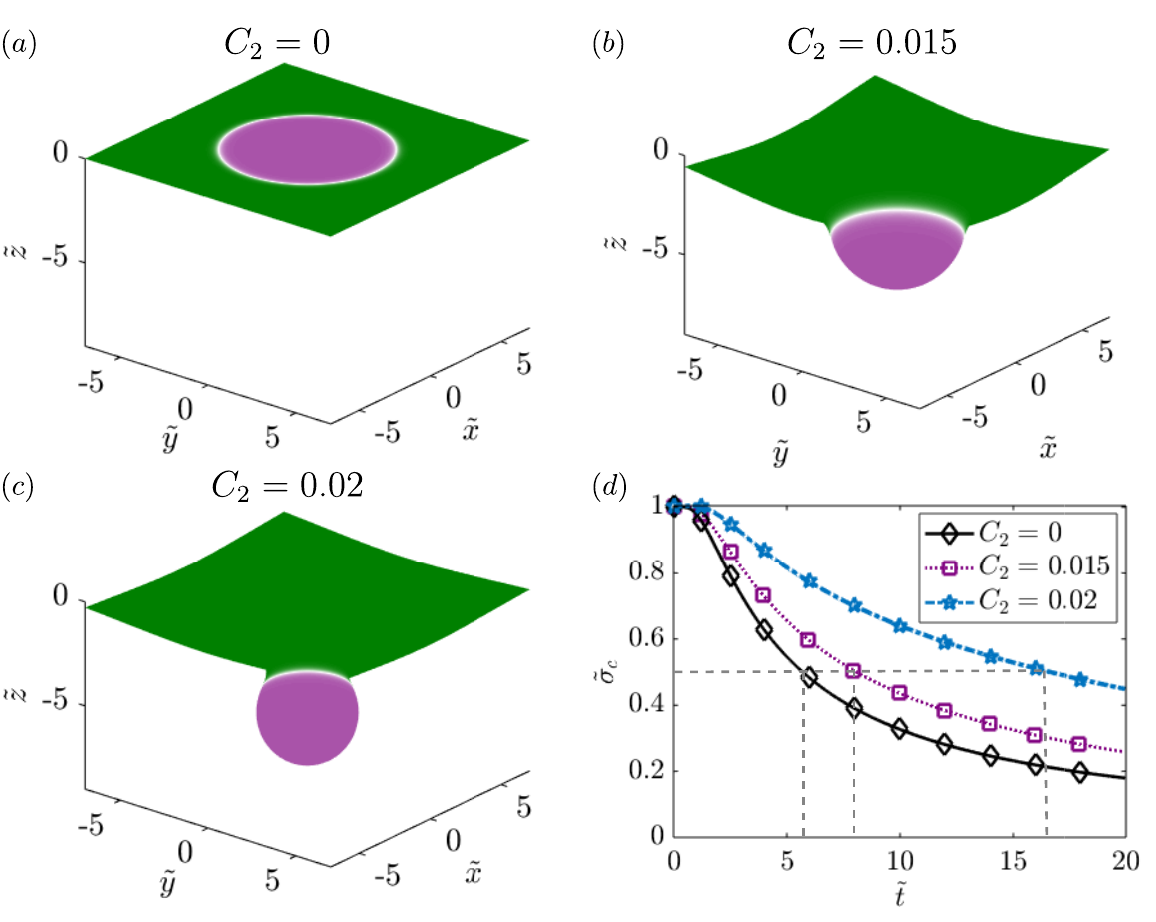}
    \caption{Dependence of surface curvature $C_2$ on the diffusion of the protein that does not induce curvature ($\ell_1=0$). Initial protein distribution on: ($a$) a flat membrane ($C_2=0$), (\textit{b}) a membrane with preexisting curvature $C_2=0.015\,$nm$^{-1}$, (\textit{c}) a membrane with preexisting curvature $C_2=0.02\,$nm$^{-1}$. (\textit{d}) Temporal dynamics of protein density at the center of the membrane for the three different configurations. The dashed lines highlight the time it takes for $\tilde{\sigma}_c$ to decrease from $1$ to $0.5$.}
    \label{fig:axi_cur_dep1}
\end{figure}

%\noindent \paragraph{\textbf{Dependence of surface curvature on protein diffusion}}\\
We first investigate the diffusion of proteins on curved surfaces by simulating the scenario where there are two types of proteins on the membrane: the first protein does not induce any curvature ($C_1=0$) but can diffuse along the membrane ($D_1=0.1\,\mu$m$^2\,$s$^{-1}$), whereas the second protein is curvature-inducing (with spontaneous curvature $C_2=0.02\,$nm$^{-1}$) 
but immobile ($D_2=0$). 
%Using two families of proteins allows us to decouple membrane shape from diffusion, and results for this case are shown in figure~\ref{fig:axi_cur_dep1}.  
Figure~\ref{fig:axi_cur_dep1}(\textit{a-c}) shows the initial shapes of the three surfaces for increasing values of the spontaneous curvature. 
\Cref{fig:axi_cur_dep1}(\textit{a}) shows the case of a flat membrane ($C_2=0$) and captures diffusion of a protein with no spontaneous curvature ($\ell=0$) similar to Fickian diffusion on a flat plane.
When the membrane is moderately curved in figure~\ref{fig:axi_cur_dep1}(\textit{b}) or heavily curved in figure~\ref{fig:axi_cur_dep1}(\textit{c}), diffusion from the center of the membrane takes a longer time compared to Fickian diffusion on a flat plane (compare purple and blue lines with the black line in  figure~\ref{fig:axi_cur_dep1}(\textit{d})). 
The time required for proteins to diffuse away from the center to the flat regions of the membrane increases with the preexisting curvature.
We compared the time taken for $\tilde{\sigma}_c$ to decrease from $1$ to $0.5$ and find that it increases nonlinearly with $C_2$,  as shown by the dashed lines in  figure~\ref{fig:axi_cur_dep1}(\textit{d}). 
This result clearly shows that the curvature of the surface alters the timescale of surface diffusion in a nonlinear fashion.

%\todoam[inline]{I think for preexisting curvature the shape of the membrane is playing key role to increase the diffusion time. We can compare this situation with dimension reduction. For example when it looks like a axisymmetric diffusion in a cylinder and the situation is similar to 1D diffusion. The effective diffusion timescale is $\frac{L^2}{2D}$ compared to the 2D case where diffusion timescale becomes $\frac{L^2}{4D}.$ }
%\todopr{6-13-2020 -- Where is this result? I also don't understand the reference to a cylinder.}

\begin{figure}
    \centering
    \includegraphics[width=\textwidth]{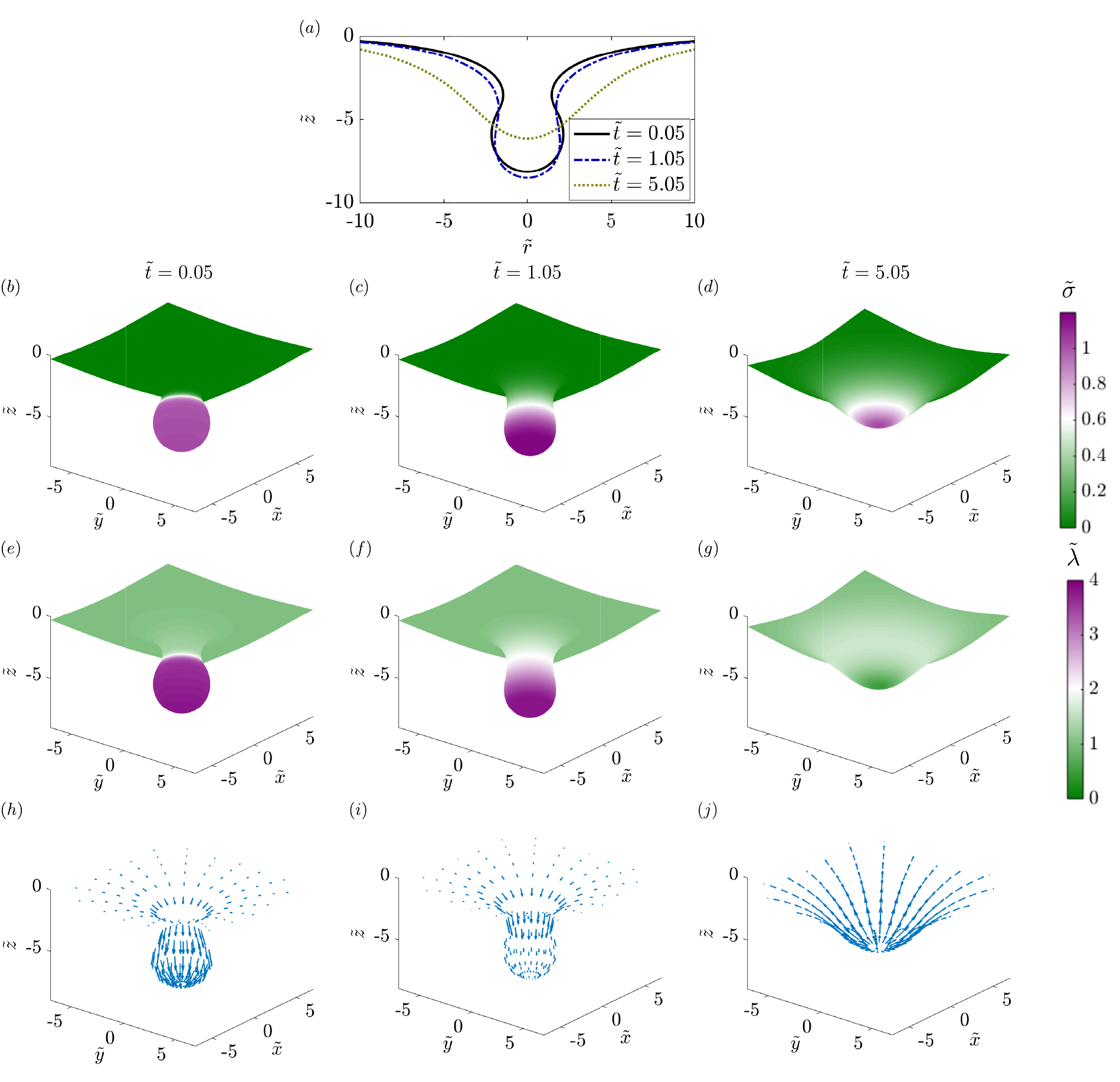}
    \caption{Dynamics of the evolution of membrane shape, protein distribution, membrane tension, and tangential velocity field at three different times. (\textit{a}) Superposed membrane shapes at three different times. ($b$-$d$) Distributions of membrane protein density are shown at dimensionless times  $0.05$, $1.05$, and $5.05$. ($e$-$g$) Distributions of membrane tension at the same non-dimensional times. ($h$-$j$)  Tangential velocity fields shown at the same non-dimensional times. Arrows are scaled according to tangential velocity magnitude, with a maximum dimensionless velocity of $5\times10^{-2}$.}
 \label{fig:axi_sym_st}
\end{figure}

Next, we simulated the full coupled system where the same protein protein induces spontaneous curvature and is free to diffuse in the plane of the membrane.  
Figure~\ref{fig:axi_sym_st} tracks the evolution of the membrane shape and protein distribution as the initial aggregate of curvature-inducing protein diffuses  over time.
At the start of the simulation, the membrane forms an $\Omega$-shaped bud with a narrow neck (figure~\ref{fig:axi_sym_st}(\textit{b})). 
The equilibrium solution of this system is a flat plane with uniform protein distribution. 
Upon initiation of the simulation, the membrane neck widens and forms a U-shaped neck. 
This widening of the neck is accompanied by a brief increase in the height of the tip of the membrane and a brief accumulation of proteins towards the bud (figure~\ref{fig:axi_sym_st}(\textit{b,c})).
Once the U-shaped neck is formed, the direction of transport reverses and proteins diffuse rapidly away from the center of the bud with a corresponding flattening of the membrane (figure~\ref{fig:axi_sym_st}\textit{d}).
The value of the membrane tension, which is initially larger at the center, eventually reduces to its boundary value as the protein density  becomes uniform (figure~\ref{fig:axi_sym_st}(\textit{e,f,g})). 
The flow profile follows the membrane deformation and the protein distribution over time. Initially, the tangential velocity is directed towards the center causing advection of the protein towards the tip (compare figure~\ref{fig:axi_sym_st}($h$) and ($i$)). 
At later times, as the protein diffuses out and the membrane begins to flatten, the flow direction reverses direction, consistent with the continuity equation (figure~\ref{fig:axi_sym_st}($j$)).

\begin{figure}
    \centering
    \includegraphics[width=\textwidth]{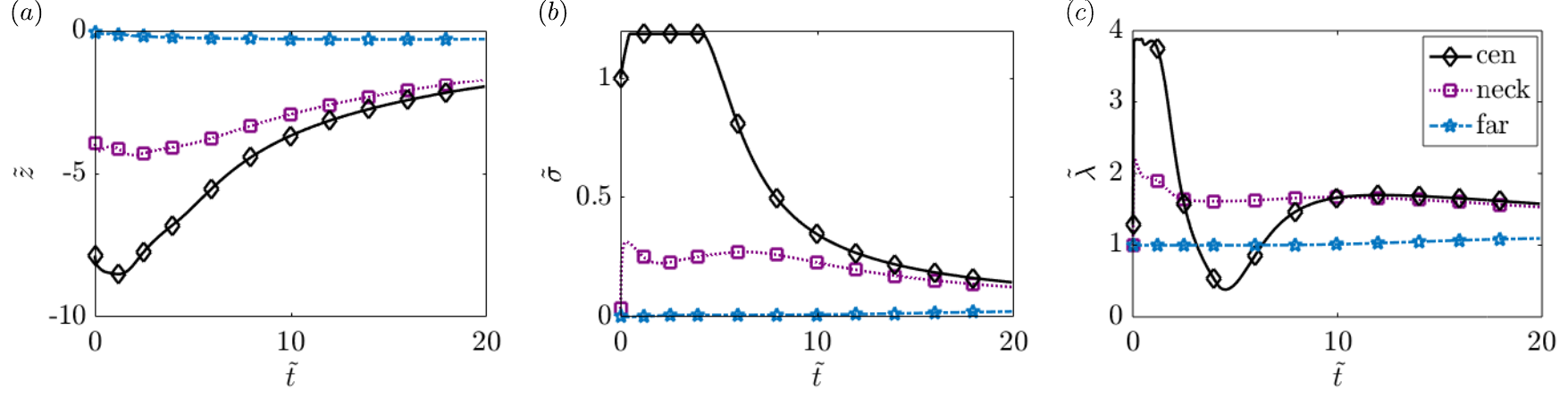}
    \caption{Temporal evolution of: ($a$) vertical displacement, ($b$) protein density and ($c$) membrane tension at three different locations: center of the membrane (cen), neck of the bud (neck) (dimensionless arclength distance from the center is $4.2$), and far from the bud (far) (dimensionless arclength distance from the center is $14.3$). } 
    \label{fig:axi_prop}
\end{figure}

\Cref{fig:axi_prop} shows the change in the displacement (\textit{a}), protein density (\textit{b}), and membrane tension (\textit{c}) at three different locations: center of the membrane (center), neck of the bud (neck), and a location far from the bud (far). 
We observe that the displacement at the center of the membrane and at the neck first increases and then decreases consistent with the initial widening of the neck (figure~\ref{fig:axi_prop}($a$)). 
No observable change in deformation was noted far from the bud.
The protein density increases at the center before decreasing over time (figure~\ref{fig:axi_prop}($b$)). We enforced a maximum value of $1.25$ for $\tilde{\sigma}$ in the simulations in place of introducing a surface saturation density of proteins on the membrane; this can be interpreted as a simple model for protein crowding.
The protein density at the neck remains more or less uniform for a long time, consistent with the diffusion of proteins away from the bud towards the flat membrane. 
The membrane tension at the center initially decreases and then increases (figure~\ref{fig:axi_prop}($c$)). 
Recall that the membrane tension is simply the negative of the surface pressure \cite{Steigmann99,Rangamani13,Rangamani14}.
The drop in the membrane tension corresponds to the change in the direction of the viscous pressure drop that results from the change in direction of the velocity field. 
Membrane tension increases further as the contribution from the viscous component becomes weak over time and the elastic component dominates. 
This is consistent with the nature of the membrane tension for the linear Monge case (figure~\ref{fig:lamlocvar}($a$)).
%\todoam{Does this text in red explain it properly?}

\begin{figure}
    \centering
    \includegraphics[width=\textwidth]{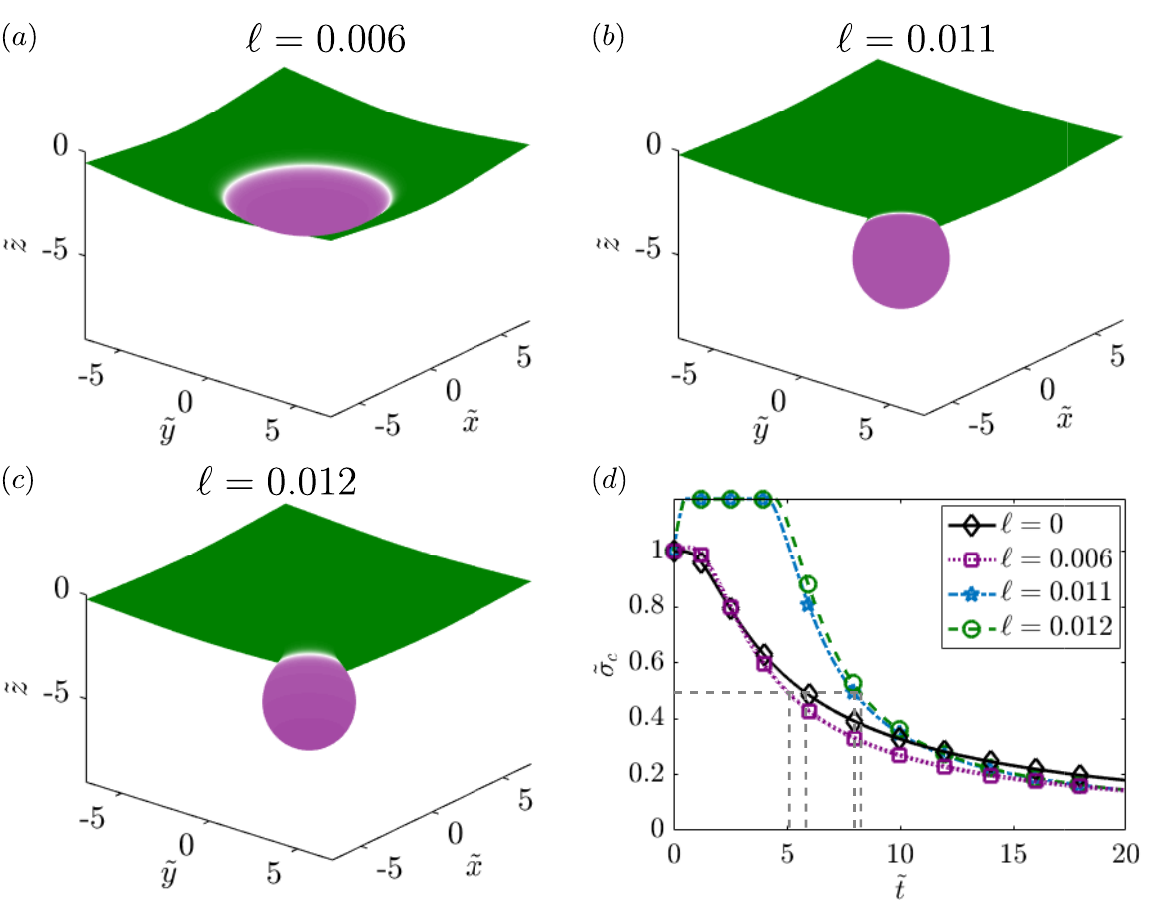}
    \caption{Coupling of protein curvature inducing effect and diffusion. Initial distribution of the protein for ($a$) $\ell=0.006\,$nm, ($b$) $\ell=0.011\,$nm, and ($c$) $\ell=0.012\,$nm.  (\textit{d}) Temporal dynamics of the density of the protein at the center of the membrane for the three different configurations. The dashed lines highlight the time it takes for $\tilde{\sigma}_c$ to decrease from $1$ to $0.5$.}
    \label{fig:axi_cur_dep}
\end{figure}

Finally, we varied the extent of the curvature induced by the protein by varying the characteristic protein size $\ell$ (figure~\ref{fig:axi_cur_dep}).
For a small value of $\ell=0.006\,$nm such that the initial curvature was a small deviation from the flat plane (figure~\ref{fig:axi_cur_dep}($a$)), protein diffusion flattens out the membrane similar to the results observed in the Monge parametrization (figure~\ref{fig:siglocvar}($a$)) and for diffusion on a flat surface (figure~\ref{fig:axi_cur_dep1}($d$)) corresponding to the case where $\ell=0$. However, increasing $\ell$ to $0.011\,$nm and $0.012\,$nm in figure~\ref{fig:axi_cur_dep}(\bit,\cit) such that the initial shape is a well-defined bud leads to altered temporal dynamics. 
We compared the time required for $\tilde{\sigma}_c$ to decrease from $1$ to $0.5$ for different values of $\ell$.
Interestingly, we find that low curvatures promote slightly faster diffusion of protein from the center of the domain as compared to flat surfaces (compare $\ell=0\,$nm to $\ell=0.006\,$nm). 
For high values of $\ell$, this time scale increases but the flattening of the membrane coupled with diffusion results in similar long-time dynamics, which is different from the case of fixed surface curvature.
Thus, we find that the coupling between membrane bending, protein diffusion, and lipid flow reveals an intricate and a somewhat counterintuitive relationship, with nonlinear dependencies between protein diffusion timescales and membrane curvature.

\section{Conclusions and discussion}
\label{sec:conclusions}
In this work, we have derived and analyzed the governing equations for the protein-induced deformation of a lipid membrane coupled with protein diffusion and in-plane viscous flow of the lipids.
The coupling between diffusion and lipid flow completes the description of the key transport phenomena involved in lipid membranes. 
We conducted simulations in 1D and 2D (linearized Monge and axisymmetry) and further quantified the relationship between membrane bending and protein diffusion.
The major conclusions from our study are that lipid flow and membrane protein diffusion, when coupled, can alter the dynamics of membrane protein distribution at different locations. 
We find that as the protein diffuses from an initial locally concentrated patch in the small deformation regime, the membrane deformation decreases and these dynamics are also related to the diffusion coefficient of proteins on the membrane.
The flow of lipids also seems to induce a separation dynamics that depends on the P\'eclet number of the system when multiple patches are present. 

In the case of buds, because of the strong coupling between protein diffusion and membrane bending, certain nonlinearities are observed. 
First, we note that the diffusion of protein at the center of a bud depends on the extent of curvature induced by the protein. 
Second, we note that in buds, proteins first tend to move towards the center of the bud to enable widening of the neck and then diffuse away from the center. 
These findings have implications for membrane flattening after fusion in cellular processes such as exocytosis \cite{hurley2010membrane} and membrane repair \cite{blazek2015plasma}.

Previously, we elaborated on the need for coupling between the viscous and elastic effects for the calculation of the Lagrange multiplier associated with the incompressibility constraint of the membrane \cite{Rangamani14, Rangamani13}. 
Here, we build on that framework to include protein diffusion. 
The coupled interaction between elasticity, diffusion, and viscous flow now fully describes the equations associated with the Lagrange multiplier $\lambda$, reinforcing its interpretation as a surface pressure \cite{Steigmann2018,Rangamani13}.
{We note that further efforts are needed in simulation technologies such that complex geometries can be simulated \cite{vasan2020mechanical,sauer2017stabilized,sahu2020arbitrary,kumar2001tri}}.

There have been many studies focused on modeling membrane-protein interactions \cite{Lipowsky12,Stachowiak10}.
Here, we show that coupling the viscous flow of lipids on the membrane is important for modulating the dynamics of the system and fully describing interfacial transport phenomena.
Future efforts will focus on adsorption of proteins from the bulk \cite{Glasmaster02,Zhdanov10} and phase separation of proteins to identify the coupling between lipid flow and chemical energies associated with these processes on an elastic membrane. 
Such theoretical developments not only have implications for our understanding of biological membranes, but also have the potential to impact curvature-driven, directed assembly in colloids and liquid crystals suspended in fluids, and particle interactions at interfaces between immiscible fluids and soft materials, enabling directed design and engineering of the next-generation of reconfigurable systems in soft matter \cite{Liu2018, Anjali19}. %\todopr{Arijit, find references from Kate Stebe (UPenn) and other related researchers, a couple will do.}.
\section*{Declaration of Interests}
The authors report no conflict of interest.
\section*{Acknowledgements}
The authors would like to thank Prof. David Steigmann for initial discussions on the model development. They would also like to thank Mr. Amaresh Sahu of UC Berkeley for his critical feedback. This work was supported in part by ONR N00014-17-1-2628  and NIH R01GM132106 to P.R., and by NSF CBET-1705377 to D.S.

\appendix

\section{Analytical justification for the flat plane as the equilibrium solution\label{app:flatsurf}}

Our 1D simulations in \S \ref{sec:1ddomain} show that at steady state the protein distribution reaches a uniform distribution while the string approaches the flat configuration. Here, we rationalize this result and prove theoretically that the flat configuration with uniform protein density is indeed an exact solution. 
To this end, we consider the arc-length parametrization and write the energy Lagrangian as 
%In both these cases, for the boundary conditions, the the analytical study to prove that flat string is the solution in the case of uniformly distributed spontaneous curvature for the stated displacement boundary conditions. To find the governing equation of displacements we have considered the arclength parameterization which makes the energy Lagrangian as
%\iffalse
%\begin{equation}
%g=\frac{d\boldsymbol{R}}{ds}.\frac{d\boldsymbol{R}}{ds}=\hat{\boldsymbol{t}}\boldsymbol{.}\hat{\boldsymbol{t}}=1.
%\end{equation}
%Now we can construct a Lagrangian subjected to local inextensibility as 
%\begin{equation}
%\mathcal{L}=\int [Jk(H-C)^2+\lambda J] ds, 
%\end{equation}
%where $J=\sqrt{g/G}$, $G(=1)$ is the first fundamental form for reference configuration, and $H$ is the curvature of the string.  We finally get
%\fi
\begin{equation}
\mathcal{L}=\int [k(H-C)^2+\lambda]\, \mathrm{d}s =\int [k(\psi_s-C)^2+\lambda]\,\mathrm{d}s.
\label{eqn:lagran}
\end{equation}
In the above relation, $\psi$ is the angle made by the string with the horizontal direction and $s$ is the arc-length. The curvature, in this case, is given by $\psi_s$. The tangential force balance reads
\begin{equation}
\label{eqn:1dlam}
2k[\psi_s-C(s)]C'(s)=\lambda_s ,
\end{equation}
which provides an equation for the string tension $\lambda$ for a given shape and spontaneous curvature.
For a uniform protein density, the spontaneous curvature is also uniform: $C(s)=C_0$, and equation (\ref{eqn:1dlam}) then implies a uniform tension $\lambda(s)=\lambda_0$ everywhere in the domain. The Lagrangian then simplifies to
\begin{equation}
\label{eqn:lagran_simp}
\mathcal{L}=\int_{-L/2}^{L/2} \left[(\psi_s-C_0)^2+\Lambda_0 \right]\,\mathrm{d}s, \quad \text{with} \quad \Lambda_0=\frac{\lambda_0}{k}.
\end{equation}
Here, we  have taken the domain to be $-L/2\le s \le L/2$, and we assume the following boundary conditions at both ends:
\begin{equation}
    y(-L/2)=y(L/2)=0, \qquad \psi(-L/2)=\psi(L/2)=0. \label{eq:appendixBC}
\end{equation}
Without loss of generality, we can consider symmetric deformations with respect to $s=0$, and seek the solution for $\psi_s$ as a Fourier cosine series of the form
\begin{equation}
 \psi_s=a_0+\sum_{n=1}^{\infty} a_n \cos \frac{2n\pi s}{L}.   
\end{equation}
Substituting this series into equation (\ref{eqn:lagran_simp}) yields 
%\iffalse
%\begin{eqnarray}
%\nonumber
 %  \mathcal{L}=\int_{-L}^{L} \left[(\psi_s-C)^2+\Lambda_0 \right] ds  \\
  % \nonumber
  %=\int_{-L}^{L} \left[(\psi_s-C)\left( a_0+\sum_{n=1}^{\infty} a_n \cos \frac{n\pi s}{L} -C\right)+\Lambda_0 \right] ds  \\
   %  \nonumber
   %=  (a_0-C) \int_{-L}^{L} \psi_s ds +  \sum_{n=1}^{\infty} a_n \int_{-L}^{L} \psi_s  \cos \frac{n\pi s}{L} ds +2L(C^2 +\Lambda_0 )\\
    %  \nonumber
   %=(a_0-C)(\psi(L)-\psi(-L))+L\sum_{n=1}^{\infty} a_n^2+2L(C^2 +\Lambda_0)\\
  %=L\sum_{n=1}^{\infty} a_n^2+2L(C^2 +\Lambda_0).
%\end{eqnarray}
%\fi
\begin{equation}
\begin{split}
   \mathcal{L}&=\int_{-L/2}^{L/2} \bigg[k\bigg( a_0+\sum_{n=1}^{\infty} a_n \cos \frac{2n\pi s}{L} -C_0\bigg)^2+\lambda_0 \bigg] \mathrm{d}s  \\
  &=\frac{L}{2}\sum_{n=1}^{\infty} k a_n^2+L(k C_0^2 +\lambda_0).
\end{split}
\end{equation}
We find that $\mathcal{L}$ is independent of $a_0$. Minimizing $\mathcal{L}$ with respect to the Fourier coefficients leads to $a_i=0$ for $i\neq0$.  We therefore find that $\phi_s=a_0$, which integrates to $\psi=a_0s+b$. Using the boundary conditions (\ref{eq:appendixBC}), we obtain
\begin{equation}
    \psi(s)=0,
\end{equation}
which indicates that the flat configuration is the equilibrium solution in this case. 

\section{Validation of algorithm for pressure-Poisson equation\label{app:numvalid}}

\begin{figure}
    \centering
    \includegraphics[width=\textwidth]{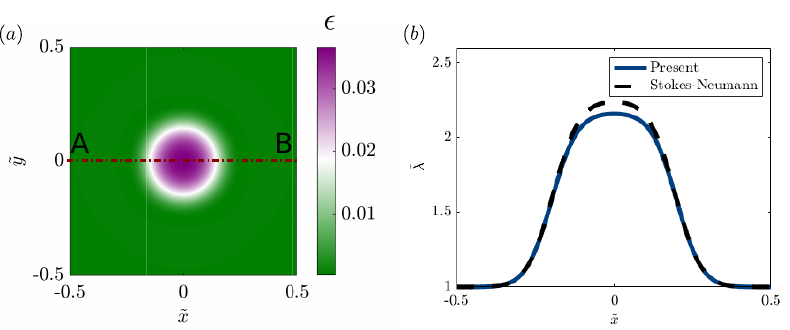}
    \caption{Comparison between membrane tension calculated using the present model and a Stokes-Neumann formulation \cite{glowinski2005numerical}. ($a$) Relative error $\epsilon=(\lambda-\lambda_{SN})/\lambda_{SN}$ in the membrane tension  for the case of single patch of protein (figure~\ref{fig:schpatch}(\textit{a})) at time $\tilde{t}=5 \times 10^{-3}$, ($b$) Membrane tension distribution along line AB shown in (a) for the present model and Stokes-Neumann solution. }
    \label{fig:validation}
\end{figure}

 In \S \ref{sec:smalldef}, we solved the coupled membrane tension and velocity for the case of linear Monge by solving the pressure-Poisson equation with the help of the integral representation of equation (\ref{eqn:int_rep_lam}). Here, we present a validation of this method and compare the result with the Stokes-Neumann system \cite{glowinski2005numerical}. Recall the governing equations for the fluid flow in the present case:
\begin{equation}
\label{eqn:gov_fluid_flow_cont}
\tilde{\boldsymbol{\nabla}}\boldsymbol{\cdot} \tilde{\boldsymbol{v}}=2\tilde{w} \tilde{H},
\end{equation}
\begin{equation}
\begin{split}
\label{eqn:gov_fluid_flow_tang}
\tilde{\boldsymbol{\nabla}} \tilde{\lambda} +\tilde{\nabla}^2 \tilde{\boldsymbol{v}}+\tilde{\boldsymbol{\nabla}} (\tilde{\boldsymbol{\nabla}}\boldsymbol{\cdot} \tilde{\boldsymbol{v}})-4\tilde{w}\tilde{\boldsymbol{\nabla}}\tilde{H}-2\tilde{\boldsymbol{\nabla}}\tilde{w}\boldsymbol{:}\tilde{\boldsymbol{\nabla}}\tilde{\boldsymbol{\nabla}}\tilde{z}&=\\
\tilde{\boldsymbol{\nabla}} \tilde{\sigma} \bigg[\frac{{\vphantom{A}\smash{2 \hat{C} \hat{B}}}}{\hat{T}} \tilde{\nabla}^2 \tilde{z} -\frac{4 \hat{C}^2 \hat{B}^2}{\hat{T}} \tilde{\sigma} &-\frac{2\hat{C}}{\hat{T}} \log\Big(\frac{\tilde{\sigma}}{\tilde{\sigma}_s} \Big) \bigg].
\end{split}
\end{equation}
The velocity field $\boldsymbol{v}$ can be written as a Helmholtz decomposition:
\begin{equation}
\begin{split}
\label{eqn:helmholtz}
\tilde{\boldsymbol{v}}&=\tilde{\boldsymbol{\nabla}}\phi + \tilde{\boldsymbol{\nabla}} \times \tilde{\boldsymbol{\zeta}}=\tilde{\boldsymbol{v}}_d+\tilde{\boldsymbol{u}}, 
\end{split}
\end{equation} 
where $\tilde{\boldsymbol{v}}_d$ is the curl-free and $\tilde{\boldsymbol{u}}$ is divergence-free. In particular, the continuity equation (\ref{eqn:gov_fluid_flow_cont}) becomes $\tilde{\boldsymbol{\nabla}}\boldsymbol{\cdot}\tilde{\boldsymbol{v}}_d=2 \tilde{w} \tilde{H}$. Now, substituting $\tilde{\boldsymbol{v}}=\tilde{\boldsymbol{u}}+\tilde{\boldsymbol{v}}_d$ into the governing equations (\ref{eqn:gov_fluid_flow_cont})--(\ref{eqn:gov_fluid_flow_tang}) for the fluid flow yields the modified system of equations:
\begin{equation}
\label{eqn:gov_mod_cont}
\tilde{\boldsymbol{\nabla}}\boldsymbol{\cdot}\tilde{\boldsymbol{u}}=0,
\end{equation}
\begin{equation}
\label{eqn:gov_mod}
\tilde{\boldsymbol{\nabla}} \tilde{\lambda} +\tilde{\nabla}^2 \tilde{\boldsymbol{u}} +\tilde{\boldsymbol{f}}=0,
\end{equation}
where,
\begin{equation}
\begin{split}
    \tilde{\boldsymbol{f}}=2\tilde{\boldsymbol{\nabla}} (2 \tilde{w} \tilde{H})-4\tilde{w}\tilde{\boldsymbol{\nabla}}\tilde{H}-2\tilde{\boldsymbol{\nabla}}\tilde{w}\boldsymbol{:}\tilde{\boldsymbol{\nabla}}\tilde{\boldsymbol{\nabla}}\tilde{z}&\\
-\tilde{\boldsymbol{\nabla}} \tilde{\sigma} \bigg[\frac{{\vphantom{A}\smash{2 \hat{C} \hat{B}}}}{\hat{T}} \tilde{\nabla}^2 \tilde{z} -\frac{4 \hat{C}^2 \hat{B}^2}{\hat{T}} \tilde{\sigma} &-\frac{2\hat{C}}{\hat{T}} \log\Big(\frac{\tilde{\sigma}}{\tilde{\sigma}_s} \Big) \bigg].
\end{split}
\end{equation}
Equations (\ref{eqn:gov_mod})--(\ref{eqn:gov_mod_cont}) constitute a non-homogeneneous Stokes problem with body force $\tilde{\boldsymbol{f}}$. We solve it here with boundary conditions $\tilde{\boldsymbol{u}}_{\infty}\rightarrow \boldsymbol{0}$ and $\tilde{{\lambda}}_{\infty}\rightarrow 1$ at infinity. In that case, the velocity and pressure are simply obtained using the boundary integral equations \cite{pozrikidis1992boundary} 
\begin{equation}
    \tilde{\boldsymbol{u}}(\tilde{\boldsymbol{x}})=\int_{\Omega} \boldsymbol{\mathcal{G}} (\tilde{\boldsymbol{x}}-\tilde{\boldsymbol{x}}_0)\boldsymbol{\cdot} \boldsymbol{f}(\boldsymbol{x}_0)\, \mathrm{d}A(\tilde{\boldsymbol{x}}_0),
\end{equation}
\begin{equation}
    \tilde{\lambda}(\tilde{\boldsymbol{x}})=1+\int_{\Omega} \boldsymbol{{\Pi}} (\tilde{\boldsymbol{x}}-\tilde{\boldsymbol{x}}_0)\boldsymbol{\cdot} \boldsymbol{f}(\boldsymbol{x}_0)\, \mathrm{d}A(\tilde{\boldsymbol{x}}_0),
\end{equation}
where $\boldsymbol{\mathcal{G}}$ and $\boldsymbol{\Pi}$ are the velocity and pressure Green's functions for two-dimensional Stokes flow and are given by:
\begin{equation}
\begin{split}
\boldsymbol{\mathcal{G}}(\tilde{\boldsymbol{x}})&=\frac{1}{4\pi} \left(\frac{\tilde{\boldsymbol{x}} \tilde{\boldsymbol{x}}}{|\tilde{\boldsymbol{x}}|^2} -\boldsymbol{I}\log |\tilde{\boldsymbol{x}}|\right),\\
    \boldsymbol{\Pi} (\tilde{\boldsymbol{x}})&=-\frac{1}{2\pi}\frac{\tilde{\boldsymbol{x}}}{|\tilde{\boldsymbol{x}}|^2}.
\end{split}
\end{equation}
Figure~\ref{fig:validation} compares the membrane tension profile $\tilde{{\lambda}}$ obtained in figure \ref{fig:schpatch}($\textit{a}$) for a single patch with the solution obtained using the Stokes-Neumann formalism. We find that the relative error is well below $4\%$ everywhere in the domain ($\textit{a}$). The two membrane tension profiles overlap over most of the domain except for a small deviation near the center ($\textit{b}$).    

\bibliographystyle{unsrt}  
\bibliography{reference}

\end{document}

%% file: main.bbl
\begin{thebibliography}{10}

\bibitem{alberts02}
B.~Alberts, A.~Johnson, J.~Lewis, M.~Raff, K.~Roberts, and P.~Walter.
\newblock {\em Molecular Biology of the Cell}.
\newblock Garland Science, 1985.

\bibitem{Harayama18}
T.~Harayama and H.~Riezman.
\newblock Understanding the diversity of membrane lipid composition.
\newblock {\em Nat. Rev. Mol. Cell Biol. Cell Biol.}, 19:281--296, 2018.

\bibitem{Bassereau18}
{Bassereau \textit{et al}.}
\newblock The 2018 biomembrane curvature and remodeling roadmap.
\newblock {\em J. Phys. D Appl. Phys.}, 51:343001, 2018.

\bibitem{Singer74}
S.~Singer.
\newblock The molecular organization of membranes.
\newblock {\em Annu. Rev. Biochem.}, 43:805--833, 1974.

\bibitem{Sackmann86}
E.~Sackmann, H.~P. Duwe, and H.~Engelhardt.
\newblock Membrane bending elasticity and its role for shape fluctuations and
  shape transformations of cells and vesicles.
\newblock {\em Faraday Discuss. Chem. Soc.}, 81:281--290, 1986.

\bibitem{Lipowsky91}
R.~Lipowsky.
\newblock The conformation of membranes.
\newblock {\em Nature}, 349:475--481, 1991.

\bibitem{Julicher93}
F.~J\"ulicher and R.~Lipowsky.
\newblock Domain-induced budding of vesicles.
\newblock {\em Phys. Rev. Lett.}, 70:2964--2967, 1993.

\bibitem{Zimmerberg06}
J.~Zimmerberg and M.~M. Kozlov.
\newblock How proteins produce cellular membrane curvature.
\newblock {\em Nat. Rev. Mol. Cell Biol.}, 7:9--19, 2006.

\bibitem{Hassinger17}
J.~E. Hassinger, G.~Oster, D.~G. Drubin, and P.~Rangamani.
\newblock Design principles for robust vesiculation in clathrin-mediated
  endocytosis.
\newblock {\em Proc. Natl. Acad. Sci.}, 114:1118--1127, 2017.

\bibitem{Antonny11}
B.~Antonny.
\newblock Mechanisms of membrane curvature sensing.
\newblock {\em Annu. Rev. Biochem.}, 80:101--123, 2011.

\bibitem{McMahon05}
H.~T. McMahon and J.~L. Gallop.
\newblock Membrane curvature and mechanisms of dynamic cell membrane
  remodeling.
\newblock {\em Nature}, 438:590--596, 2005.

\bibitem{Stachowiak10}
J.~C. Stachowiak, C.~C. Hayden, and D.~Y. Saski.
\newblock Steric confinement of proteins on lipid membranes can drive curvature
  and tubulation.
\newblock {\em Proc. Natl. Acad. Sci.}, 107:7781--7786, 2010.

\bibitem{Reynwar07}
B.~J. Reynwar, G.~Illya, V.~A. Harmandaris, M.~M. M\"uller, K.~Kremer, and
  M.~Deserno.
\newblock Aggregation and vesiculation of membrane proteins by
  curvature-mediated interactions.
\newblock {\em Nature}, 447:461--464, 2007.

\bibitem{Mukherjee00}
S.~Mujherjee and E.~R. Maxfield.
\newblock Role of membrane organization and membrane domains in endocytic lipid
  trafficking.
\newblock {\em Traffic.}, 1:203--211, 2000.

\bibitem{Gruenberg01}
J.~Gruenberg.
\newblock The endocytic pathway: A mosaic of domains.
\newblock {\em Nat. Rev. Mol. Cell Biol.}, 2:721--730, 2001.

\bibitem{Leibler87}
S.~Leibler and D.~Andelman.
\newblock Ordered and curved meso-structures in membranes and amphiphilic
  films.
\newblock {\em J. Phys. France.}, 48:2013, 1987.

\bibitem{seifert1993curvature}
U.~Seifert.
\newblock Curvature-induced lateral phase segregation in two-component
  vesicles.
\newblock {\em Phys. Rev. Lett.}, 70:1335, 1993.

\bibitem{Kahya04}
N.~Kahya, D.~Scherfeld, K.~Bacia, and P.~Schwille.
\newblock Lipid domain formation and dynamics in giant unilamellar vesicles
  explored by fluorescence correlation spectroscopy.
\newblock {\em J. Struct. Biol.}, 147:77--89, 2004.

\bibitem{Tran-son-tay84}
R.~Tran-Son-Tay, S.~P. Sutera, and P.~R. Rao.
\newblock Determination of red blood cell membrane viscosity from rheoscopic
  observations of tank-treading motion.
\newblock {\em Biophys. J.}, 46:1335--1338, 1984.

\bibitem{Noguchi04}
H.~Noguchi and G.~Gompper.
\newblock Fluid vesicles with viscous membranes in shear flow.
\newblock {\em Phys. Rev. Lett.}, 93:258102, 2004.

\bibitem{Baumgart03}
T.~Baumgart, S.~T. Hess, and W.~W. Webb.
\newblock Imaging coexisting fluid domains in biomembrane models coupling
  curvature and line tension.
\newblock {\em Nature}, 425:821--824, 2003.

\bibitem{Horner09}
A.~Horner, Y.~N. Antonenko, and P.~Pohl.
\newblock Coupled diffusion of peripherally bound peptides along the outer and
  inner membrane leaflets.
\newblock {\em Biophys. J.}, 96:2689--2695, 2009.

\bibitem{Stachowiak17}
W.~T. Snead, C.~C. Hayden, A.~K. Gadok, E.~M.~Lafer C.~Zhao, P.~Rangamani, and
  J.~C. Stachowiak.
\newblock Membrane fission by protein crowding.
\newblock {\em Proc. Natl. Acad. Sci.}, 114:3258--3267, 2017.

\bibitem{Lowengrub2009}
J.~S. Lowengrub, A.~Ratz, and A.~Voigt.
\newblock Phase-field modeling of the dynamics of multicomponent vesicles:
  Spinodal decomposition, coarsening, budding, and fission.
\newblock {\em Phys. Rev. E}, 79:031926, 2009.

\bibitem{Elliot2013}
C.~M. Elliott and B.~Stinner.
\newblock Computation of two-phase biomembranes with phase dependent material
  parameters using surface finite elements.
\newblock {\em Commun. Comput. Phys.}, 13:325--360, 2013.

\bibitem{Elliot2016}
C.~M. Elliott, C.~Graser, G.~Hobbs, R.~Kornhuber, and M.~A. Wolf.
\newblock A variational approach to particles in lipid membranes.
\newblock {\em Arch. Rational Mech. Anal.}, 222:1011--1075, 2016.

\bibitem{Seifert93b}
U.~Seifert and S.~A. Langer.
\newblock Viscous modes of fluid bilayer membranes.
\newblock {\em Europhys. Lett.}, 23:65--72, 1993.

\bibitem{sahu2017irreversible}
A.~Sahu, R.~A. Sauer, and K.~K. Mandadapu.
\newblock Irreversible thermodynamics of curved lipid membranes.
\newblock {\em Phys. Rev. E}, 96:042409, 2017.

\bibitem{Helfrich73}
W.~Helfrich.
\newblock Elastic properties of lipid bilayers: Theory and possible
  experiments.
\newblock {\em Z. Naturforsch. C}, 5:693--703, 1973.

\bibitem{Canham70}
P.~B. Canham.
\newblock The minimum energy of bending as a possible explanation of the
  biconcave shape of the human red blood cell.
\newblock {\em J. Theoret. Biol.}, 26:61--81, 1970.

\bibitem{Jenkins77}
J.~T. Jenkins.
\newblock The equation of mechanical equilibrium of a model membrane.
\newblock {\em SIAM J. Appl. Math.}, 4:693--703, 1977.

\bibitem{Steigmann99}
D.~J. Steigmann.
\newblock Fluid films with curvature elasticity.
\newblock {\em Arch. Rat. Mech.}, 150:257--272, 1999.

\bibitem{Arroyo09}
M.~Arroyo and A.~DeSimone.
\newblock Relaxation dynamics of fluid membranes.
\newblock {\em Phys. Rev. E}, 79:031915, 2009.

\bibitem{Arroyo12}
M.~Rahimi and M.~Arroyo.
\newblock Shape dynamics, lipid hydrodynamics, and complex viscoelasdticity of
  bilayer membranes.
\newblock {\em Phys. Rev. E}, 86:011932, 2012.

\bibitem{Rangamani13}
P.~Rangamani, A.~Agrawal, K.~Mandadapu, G.~Oster, and D.~Steigmann.
\newblock Interaction between surface shape and intra-surface viscous flow on
  lipid membranes.
\newblock {\em Biomech. Model. Mechanobiol.}, 12:833--845, 2013.

\bibitem{Arroyo19}
C.~Tozzi, N.~Walani, and M.~Arroyo.
\newblock Out-of-equilibrium mechanochemistry and self-organization of fluid
  membranes interacting with curved proteins.
\newblock {\em New J. Phys.}, 21:093004, 2019.

\bibitem{Scriven60}
L.~E. Scriven.
\newblock Dynamics of a fluid surface.
\newblock {\em Chem. Eng. Sci.}, 12:98--108, 1960.

\bibitem{Rangamani14}
P.~Rangamani, K.~Mandadapu, and G.~Oster.
\newblock Protein induced membrane curvature alters local membrane tension.
\newblock {\em Biophys. J.}, 107:751--762, 2014.

\bibitem{Bahmani15}
F.~Bahmani, J.~Christenson, and P.~Rangamani.
\newblock Analysis of lipid flow on minimal surfaces.
\newblock {\em Continuum Mech. Thermodyn.}, 28:503--513, 2016.

\bibitem{Iglic05}
A.~Igli\v{c}, B.~Babnik, U.~Gimsa, and V.~Kralj-Igli\v{c}.
\newblock On the role of membrane anisotropy in the beading transition of
  undulated tubular membrane structures.
\newblock {\em J. Phys. A}, 40:8527--8536, 2005.

\bibitem{Iglic05a}
V.~Kralj-Igli\v{c}, H.~Hagerstand, P.~Veranic, K.~Jezernik, B.~Babnik, D.~R.
  Gauger, and A.~Igli\v{c}.
\newblock Amphiphile-induced tubular budding of the bilayer membrane.
\newblock {\em Eur. Biophys. J.}, 34:1066--1070, 2005.

\bibitem{Gov07}
A.~Veksler and N.~S. Gov.
\newblock Phase transitions of the coupled membrane-cytoskeleton modify
  cellular shape.
\newblock {\em Biophys. J.}, 93:3798--3810, 2007.

\bibitem{Ramaswamy99}
S.~Ramaswamy, J.~Toner, and J.~Prost.
\newblock Nonequilibrium fluctuations, travelling waves, and instabilities in
  active membranes.
\newblock {\em Phys. Rev. Lett.}, 84:3494--3497, 2005.

\bibitem{Gozdz11}
W.~T. Gozdz.
\newblock Shape transformation of lipid vesicle induced by diffusing
  macromolecules.
\newblock {\em J. Chem. Phys.}, 134:371--379, 2011.

\bibitem{Agrawal11}
D.~Steigmann and A.~Agrawal.
\newblock A model for surface diffusion of trans-membrane protein on lipid
  bilayers.
\newblock {\em Z. Angew. Math. Phys.}, 62:449--563, 2011.

\bibitem{Arroyo18}
M.~Arroyo, N.~Walani, A.~Torres-Sanchez, and D.~Kaurin.
\newblock Onsager's variational principle in soft matter: {I}ntroduction and
  application to the dynamics of adsorption of proteins onto fluid membranes.
\newblock In D.~J. Steigmann, editor, {\em The Role of Mechanics in the Study
  of Lipid Bilayers}, pages 1--53. Springer, Cham, Switzerland, 2018.

\bibitem{Arroyo19jfm}
A.~Torres-Sanchez, D.~Millan, and M.~Arroyo.
\newblock Modelling fluid deformable surfaces with an emphasis on biological
  interfaces.
\newblock {\em J. Fluid Mech.}, 872:218--271, 2019.

\bibitem{Steigmann2018}
D.~J. Steigmann.
\newblock Mechanics and physics of lipid bilayers.
\newblock In D.~J. Steigmann, editor, {\em The Role of Mechanics in the Study
  of Lipid Bilayers}, pages 1--61. Springer, Cham, Switzerland, 2018.

\bibitem{Sokolnikoff1964}
I.~S. Sokolnikoff.
\newblock {\em Tensor Analysis: Theory and Applications}.
\newblock Wiley, 1951.

\bibitem{Kreyszig1968}
E.~Kreyszig.
\newblock {\em Advanced Engineering Mathematics}.
\newblock Wiley, 1968.

\bibitem{Aris89}
R.~Aris.
\newblock {\em Vectors, Tensors and Basic Equation of Fluid Mechanics}.
\newblock Dover, New York, 1989.

\bibitem{Steigmann03}
D.~Steigmann, E.~Baesu, R.~E. Rudd, J.~Belak, and M.~McEleresh.
\newblock On the variational theory of cell-membrane equilibria.
\newblock {\em Interfaces Free Bound.}, 5:357--366, 2003.

\bibitem{Alimohamadi18}
H.~Alimohamadi and P.~Rangamani.
\newblock Modeling membrane curvature generation due to membrane– protein
  interactions.
\newblock {\em Biomolecules}, 8:120--147, 2018.

\bibitem{Gov2018}
N.~S. Gov.
\newblock Guided by curvature: shaping cells by coupling curved membrane
  proteins and cytoskeletal forces.
\newblock {\em Philos. Trans. R. Soc. Lond. B Biol. Sci.}, 373:20170115, May
  2018.

\bibitem{Julicher17}
G.~Salbreux and F.~{J\"ulicher}.
\newblock Mechanics of active surfaces.
\newblock {\em Phys. Rev. E}, 96:032404, 2017.

\bibitem{Mietke19}
A.~Mietke, F.~Julicher, and I.~F. Sbalzarini.
\newblock Self-organized shape dynamics of active surfaces.
\newblock {\em Proc. Natl. Acad. Sci.}, 116:29--34, 2019.

\bibitem{mikucki2017curvature}
M.~Mikucki and Y.~C. Zhou.
\newblock Curvature-driven molecular flow on membrane surface.
\newblock {\em SIAM J. Appl. Math.}, 77:1587--1605, 2017.

\bibitem{gera2017cahn}
P.~Gera and D.~Salac.
\newblock Cahn--{Hilliard} on surfaces: A numerical study.
\newblock {\em Appl. Math. Lett.}, 73:56--61, 2017.

\bibitem{Do76}
C.~Do.
\newblock {\em Differential Geometry of Curves and Surfaces}.
\newblock Prentice Hall, 1976.

\bibitem{chabanon2018}
M.~Chabanon and P.~Rangamani.
\newblock Gaussian curvature directs the distribution of spontaneous curvature
  on bilayer membrane necks.
\newblock {\em Soft Matter}, 12:2281--2294, 2018.

\bibitem{chabanon2019}
M.~Chabanon and P.~Rangamani.
\newblock Geometric coupling of helicoidal ramps and curvature-inducing
  proteins in organelle membranes.
\newblock {\em J. R. Soc. Interface.}, 16:20190354, 2019.

\bibitem{gozdz1998composition}
W.~T. G{\'o}{\'z}d{\'z} and G.~Gompper.
\newblock Composition-driven shape transformations of membranes of complex
  topology.
\newblock {\em Phys. Rev. Lett.}, 80:4213, 1998.

\bibitem{campelo2007model}
F.~Campelo and A.~Hern{\'a}ndez-Machado.
\newblock Model for curvature-driven pearling instability in membranes.
\newblock {\em Phys. Rev. Lett.}, 99:088101, 2007.

\bibitem{ouyang1989bending}
O.-Y. Zhong-Can and W.~Helfrich.
\newblock Bending energy of vesicle membranes: General expressions for the
  first, second, and third variation of the shape energy and applications to
  spheres and cylinders.
\newblock {\em Phys. Rev. A}, 39:5280, 1989.

\bibitem{glowinski2005numerical}
R.~Glowinski, T.-W. Pan, V.~L.~H. Juarez, and E.~Dean.
\newblock Numerical methods for the simulation of incompressible viscous flow:
  An introduction.
\newblock In {\em Multidisciplinary Methods for Analysis Optimization and
  Control of Complex Systems}, pages 49--175. Springer, 2005.

\bibitem{haucke2018membrane}
V.~Haucke and M.~M. Kozlov.
\newblock Membrane remodeling in clathrin-mediated endocytosis.
\newblock {\em J. Cell Sci.}, 131:jcs216812, 2018.

\bibitem{dannhauser2012reconstitution}
P.~N. Dannhauser and E.~J. Ungewickell.
\newblock Reconstitution of clathrin-coated bud and vesicle formation with
  minimal components.
\newblock {\em Nature Cell Biol.}, 14:634--639, 2012.

\bibitem{Rangamani20}
P.~Rangamani, A.~Behzadan, and M.~Holst.
\newblock Local sensitivity analysis of the `membrane shape equation' derived
  from {Helfrich} energy.
\newblock {\em arxiv}, 2005:12550, 2020.

\bibitem{kozlovsky2003fission}
Y.~Kozlovsky and M.~M. Kozlov.
\newblock Membrane fission: model for intermediate structures.
\newblock {\em Biophys. J.}, 85:85--96, 2003.

\bibitem{jahn2003fusion}
R.~Jahn, T.~Lang, and T.~C. S{\"u}dhof.
\newblock Membrane fusion.
\newblock {\em Cell}, 112:519--533, 2003.

\bibitem{hurley2010membrane}
J.~H. Hurley, E.~Boura, L.-A. Carlson, and B.~R{\'o}{\.z}ycki.
\newblock Membrane budding.
\newblock {\em Cell}, 143:875--887, 2010.

\bibitem{blazek2015plasma}
A.~D. Blazek, B.~J. Paleo, and N.~Weisleder.
\newblock Plasma membrane repair: a central process for maintaining cellular
  homeostasis.
\newblock {\em Physiology}, 30:438--448, 2015.

\bibitem{vasan2020mechanical}
R.~Vasan, S.~Rudraraju, M.~Akamatsu, K.~Garikipati, and P.~Rangamani.
\newblock A mechanical model reveals that non-axisymmetric buckling lowers the
  energy barrier associated with membrane neck constriction.
\newblock {\em Soft Matter}, 16:784--797, 2020.

\bibitem{sauer2017stabilized}
R.~A. Sauer, T.~X. Duong, K.~K. Mandadapu, and D.~J. Steigmann.
\newblock A stabilized finite element formulation for liquid shells and its
  application to lipid bilayers.
\newblock {\em J. Comput. Phys.}, 330:436--466, 2017.

\bibitem{sahu2020arbitrary}
A.~Sahu, Y.~A.~D. Omar, R.~A. Sauer, and K.~K. Mandadapu.
\newblock Arbitrary {Lagrangian--Eulerian} finite element method for curved and
  deforming surfaces: {I. General} theory and application to fluid interfaces.
\newblock {\em J. Comput. Phys.}, 407:109253, 2020.

\bibitem{kumar2001tri}
P.~B.~S. Kumar, G.~Gompper, and R.~Lipowsky.
\newblock Budding dynamics of multicomponent membranes.
\newblock {\em Phys. Rev. Lett.}, 86:3911, 2001.

\bibitem{Lipowsky12}
R.~Lipowsky.
\newblock Spontaneous tubulation of membranes and vesicles reveals membrane
  tension generated by spontaneous curvature.
\newblock {\em Faraday Discuss.}, 161:305--331, 2013.

\bibitem{Glasmaster02}
K.~Glasmaster, C.~Larsson, F.~Hook, and B.~Kasemo.
\newblock Protein adsorption on supported phospholipid bilayer.
\newblock {\em J. Colloid Interface Sci.}, 246:40--47, 2002.

\bibitem{Zhdanov10}
V.~P. Zhdanov and B.~Kasemo.
\newblock Adsorption of proteins on a lipid bilayer.
\newblock {\em Euro. Biophys. J.}, 39:1477--1482, 2010.

\bibitem{Liu2018}
I.~B. Liu, N.~Sharifi-Mood, and K.~J. Stebe.
\newblock Capillary assembly of colloids: Interactions on planar and curved
  interfaces.
\newblock {\em Annu. Rev. Condens. Matter Phys.}, 9:283--305, March 2018.

\bibitem{Anjali19}
T.~G. Anjali and M.~G. Basavaraj.
\newblock Shape-anisotropic colloids at interfaces.
\newblock {\em Langmuir}, 35:3--20, 2019.

\bibitem{pozrikidis1992boundary}
C.~Pozrikidis.
\newblock {\em Boundary Integral and Singularity Methods for Linearized Viscous
  Flow}.
\newblock Cambridge University Press, 1992.

\end{thebibliography}
